# Solar Daylighting to Offset LED Lighting in Vertical Farming: A Techno-Economic Study of Light Pipes

Francesco Ceccanti[1]*, Aldo Bischi[1], Marco Antonelli[1], Andrea Baccioli[1]

1 Department of Energy, Systems, Territory and Construction Engineering
Università di Pisa, Pisa, Italy

*Corresponding Author: francesco.ceccanti@phd.unipi.it (F. Ceccanti)

**Abstract**

Vertical farming is a controlled-environment agriculture (CEA) approach in which crops are grown in stacked layers under regulated climate and lighting, enabling predictable production but requiring high electricity input. This study quantifies the techno-economic impact of roof-mounted daylighting in a three-tier container vertical farm using a light-pipe (LP) system that delivers sunlight to the upper tier. The optical chain, comprising a straight duct and a tilting aluminum-coated mirror within a rotating dome, was modelled in Tonatiuh to estimate crop-level photon delivery and solar gains. These outputs were coupled with a transient AGRI-Energy model to perform year-round simulations for Dubai. Tier-3 strategies were compared against a fully LED benchmark, including daylight-only operation, on/off supplementation, PWM dimming, UV–IR filtering, variable-transmittance control, and simple glazing. Ray-tracing predicted a crop-level LP optical efficiency of 45%-75%, depending on solar position, quantifying the fraction of incident daylight at the collector aperture delivered to the target growing zone. Daylight-only operation reduced the total three-tier yield by 17% and was not economically viable despite 27–29% electricity savings. Hybrid daylight–LED strategies preserved benchmark yield while reducing electricity use. PWM dimming combined with UV–IR filtering achieved the lowest specific electricity energy consumption (6.32 kWh $kg^{-1}$), 14% below the benchmark. Overall, viability remains CAPEX-limited because achievable electricity savings are insufficient to offset the added investment and thus improves mainly under high electricity and carbon-price contexts, although the LP system delivers a 15–38% lower light cost than an optical-fiber reference under identical incident daylight.

Nomenclature

| **Acronyms** | | | **Subscripts** | |
|---|---|---|---|---|
| **AHU** | Air Handling Unit | $[-]$ | **ahu** | Air Handling Unit |
| **CAPEX** | Capital Expenditures | $[\$]$ | **conv** | Convective |
| **CEA** | Controlled Environment Agriculture | $[-]$ | **$CO_2$** | Carbon Dioxide |
| **COP** | Coefficient of Performance | $[kW_{th}\, kW_{el}^{-1}]$ | **diff** | Diffuse |
| **DHI** | Diffuse Horizontal Irradiance | $[W\, m^{-2}]$ | **dir** | Direct |
| **DLI** | Daily Light Integral | $[\mu mol\, m^{-2}\, day^{-1}]$ | **dm** | Dry Matter |
| **DNI** | Direct Normal Irradiance | $[W\, m^{-2}]$ | **EC** | Electric-Controlled |
| **HVAC** | Heating, Ventilation and Air Conditioning | $[-]$ | **el** | electrical |
| **LAI** | Leaf Area Index | $[{m_{leaf}}^{2}\, {m_{soil}}^{-2}]$ | **env** | Envelope |

| **LC** | Light Cost | $[\$\ (\mu mol\ s^{-1})^{-1}]$ | **eva** | Evaporation |
|---|---|---|---|---|
| **LCoL** | Levelized Cost of Lettuce | $[\$\ kg^{-1}]$ | **fm** | Fresh Matter |
| **LP** | Light Pipe | $[-]$ | **H/C** | Heating & Cooling Systems |
| **LUE** | Light Use Efficiency | $[g\ \mu mol^{-1}]$ | **hum** | Humidification System |
| **PAR** | Photosynthetic Active Radiation | $[\mu mol\ m^{-2}\ s^{-1}]$ | **LED** | Lighting System |
| **PBT** | Payback Time | $[years]$ | **plant** | Single Lettuce Plant |
| **PPE** | Photosynthetic Photon Efficiency | $[\mu mol\ J^{-1}]$ | **ref** | Reference |
| **PPFD** | Photosynthetic Photon Flux Density | $[\mu mol\ m^{-2}\ s^{-1}]$ | **sol** | Solar |
| **PWM** | Pulse Width Modulation | $[-]$ | **th** | Thermal |
| **SEC** | Specific Energy Consumption | $[kWh\ kg^{-1}]$ | | |
| **SEEC** | Specific Electric Energy Consumption | $[kWh\ kg^{-1}]$ | | |
| **VF** | Vertical Farm | $[-]$ | | |
| **WUE** | Water Use Efficiency | $[g_{FM}\ L^{-1}]$ | | |
| **Symbols** | | | | |
| $\boldsymbol{A}$ | Area | $[m^2]$ | | |
| $\boldsymbol{B}$ | Auxiliary Angle | $[°]$ | | |
| $\boldsymbol{c}$ | Specific Cost | $[\$\ x^{-1}]$ | | |
| $\boldsymbol{c_p}$ | Specific Heat Capacity | $[J\ kg^{-1}\ K^{-1}]$ | | |
| $\boldsymbol{d}$ | Density | $[\$]$ | | |
| $\boldsymbol{DM}$ | Dry Matter | $[g\ m^{-2}]$ | | |
| $\boldsymbol{g}$ | Gravity Acceleration | $[m\ s^{-2}]$ | | |
| $\boldsymbol{E}$ | Energy | $[kWh]$ | | |
| $\boldsymbol{EoT}$ | Equation of Time | $[min]$ | | |
| $\boldsymbol{f}$ | $CO_2$ factor for crop growth | $[-]$ | | |
| $\boldsymbol{FM}$ | Fresh Matter | $[-]$ | | |
| $\boldsymbol{I}$ | Solar Irradiation | $[-]$ | | |
| $\boldsymbol{h}$ | Time | $[hour]$ | | |
| $\boldsymbol{k}$ | Light Extinction Coefficient | $[-]$ | | |
| $\boldsymbol{L}$ | Characteristic Length | $[m]$ | | |
| $\boldsymbol{n}$ | Day | $[day]$ | | |
| $\boldsymbol{Nu}$ | Nusselt's Number | $[-]$ | | |
| $\boldsymbol{Pr}$ | Prandtl's Number | $[-]$ | | |
| $\boldsymbol{Q}$ | Thermal Power | $[kW]$ | | |
| $\boldsymbol{Ra}$ | Rayleigh's Number | $[-]$ | | |
| $\boldsymbol{T}$ | Temperature | $[°C]$ | | |
| $\boldsymbol{t}$ | Time | $[s]$ | | |
| $\boldsymbol{U}$ | Thermal Transmittance | $[W\ m^{-2}\ K^{-1}]$ | | |
| $\boldsymbol{v}$ | Voltage | $[V]$ | | |
| $\boldsymbol{V}$ | Volume | $[m^3]$ | | |
| **Greeks** | | | | |
| $\boldsymbol{\alpha}$ | Altitude | $[°]$ | | |
| $\boldsymbol{\beta}$ | Thermal Expansion Coefficient | $[K^{-1}]$ | | |
| $\boldsymbol{\gamma}$ | Azimuth Angle | $[°]$ | | |
| $\boldsymbol{\psi}$ | Longitude | $[°]$ | | |
| $\boldsymbol{\eta}$ | Efficiency | $[-]$ | | |
| $\boldsymbol{\varphi}$ | Latitude | $[°]$ | | |
| $\boldsymbol{\delta}$ | Declination Angle | [°] | | |
| $\boldsymbol{\omega}$ | Hour Angle | [°] | | |
| $\boldsymbol{\tau}$ | Optical Transmission | [−] | | |
| $\boldsymbol{\theta}$ | Incidence Angle | [°] | | |

## 1. Introduction

In recent years, the need to produce food more sustainably has made controlled environment agriculture (CEA) an increasingly important and promising approach. Within this field, vertical farms (VFs) grow crops in stacked layers inside enclosed spaces such as buildings, shipping containers, or dedicated modules, typically using hydroponic or aeroponic systems [1]. They usually operate with recirculating nutrient solutions and electric lighting, which reduces dependence on outdoor weather and seasonal conditions. Because temperature, humidity, $CO_2$ levels, and lighting schedules can be managed with precision, vertical farming is often seen as a technological option to address urban growth, climate-related risks to conventional agriculture, and supply chain disruptions that affect the availability and quality of fresh produce [2]. At the same time, the sector has drawn strong interest from investors and policymakers, and many market reports forecast rapid expansion, even though estimates differ depending on definitions and scope. Overall, vertical farming is gaining visibility both as an agrifood business model and as an active area of research [3].

Vertical farming is often promoted for its potential gains in resource efficiency and product quality. Stacked, year-round production can deliver high yields per unit area, which is especially attractive where arable land is scarce or expensive [4]. Closed-loop irrigation can reduce water use compared with conventional systems, although savings depend on climate and system design [5]. Finally, indoor cultivation can limit pest and weather exposure, lowering pesticide needs and improving crop consistency, while proximity to consumers can shorten distribution and reduce spoilage, particularly for leafy greens. These factors underpin the role of VFs in resilient urban food strategies for high-value and short-cycle crops [6].

However, the main constraint of vertical farming is its high energy demand, which directly affects both operating costs and environmental impacts [7]. Electricity use is driven primarily by artificial lighting and by climate control, including HVAC and dehumidification, making energy efficiency a key performance indicator for both profitability and sustainability. Recent benchmarking work reports specific electricity consumption for lettuce of roughly 10 to 18 kWh per kg [8], and notes that improved technologies and tighter operational control could reduce these values over time, even if reliance on electricity remains fundamental [8]. Technoeconomic studies also show that production costs are highly sensitive to local electricity prices and external climate conditions. For example, an energy and cost case study on lettuce growth chambers found large location-dependent differences in annual energy demand, with energy costs contributing about 0.46 to 1.74 dollars per kg across selected cities [7]. This sensitivity has clear consequences, as industry reports and recent coverage link financial strain, downsizing, and business failures in parts of the indoor farming sector to persistently high or volatile energy prices.

Most of the electricity demand in vertical farming is ultimately driven by photon delivery, with artificial lighting dominating energy consumption and HVAC providing the cooling and dehumidification required to remove the associated sensible and latent heat. Indeed, the container farm designed and tested in [9], LEDs accounted for 72% of daily electricity use, HVAC for 14%, and auxiliary loads for the remaining 14%. For this reason, a key research priority is to improve light delivery efficiency and, where possible, reduce reliance on fully electric illumination by leveraging natural light [10]. A comparative study between a plant factory (fully indoor) and greenhouses illustrates this trade-off. The study reported that the plant factory achieved lower overall specific energy consumption and better resource use efficiency for land, $CO_2$, and water than the greenhouses [11]. However, the greenhouses were still found to be more economically attractive because they can exploit solar radiation at no direct energy cost. In terms of energy use, a greenhouse located in the Netherlands was reported to require about 2,100 MJ per kg of dry biomass when accounting for both natural and purchased energy inputs. Solar radiation provides the largest share of the total energy input (76%), while heating and cooling each accounted for about 12%. As a consequence, the demand for purchased electricity was much lower, on the order of 70 $kWh_e$ per kg of dry biomass. In this greenhouse case, cooling is largely managed through natural ventilation, so purchased energy is mainly associated with heating and auxiliary equipment. Under the same comparison, the plant factory required about 33% less total energy input, but about 253% more electricity (247 vs 70 $kWh_e$ per kg dry biomass), with artificial lighting accounting for around 90% of the electrical load [11].

Hence, the first objective of this paper is to investigate how electricity use can be reduced in a vertical farm by adopting daylighting strategies that decrease the amount of artificial light required. Daylighting is the deliberate use of natural light to illuminate indoor spaces through architectural and optical design [12]. Its purpose is twofold: to increase useful light for occupants or crops and to cut reliance on electric lighting. In practice, daylighting can be achieved through façade and roof openings, reflective or light redirecting surfaces, and dedicated daylight transport systems such as tubular devices (light pipes (LPs)) [12], which capture outdoor light and channel it into deeper indoor areas using highly reflective ducts and diffusers. Because daylight is delivered without electricity during operation, daylighting is widely studied as an energy efficiency measure that can reduce lighting runtimes and the associated cooling loads. For these reasons, it has also been proposed as a promising option to offset part of the lighting demand in multi-level buildings [13].

Within controlled environment agriculture, several daylighting approaches have already been explored, reflecting a clear research interest in reducing the electricity demand of indoor farming. For example, a recent study on a hybrid urban vertical greenhouse implemented a simple control system based on light sensors, LEDs, and microcontrollers, reporting a specific LED electricity consumption of 1.64 to 4.90 kWh per kg [10]. However, relying on direct sunlight entering through façades or roofs, as in conventional greenhouses, raises important challenges in urban contexts. One key limitation is external shading. Eaton et al. estimated that surrounding buildings can drastically reduce the light reaching the crops, to the point that solar radiation may cover only about 5% of the total light requirement [14]. While their analysis focused on a skyscraper farm, the constraint is likely even stronger in multi-tier vertical farms, where stacked shelves cause additional self-shading and further limit daylight penetration. To address these limitations, two studies investigated design solutions that integrate façade-based daylight with vertical farming layouts. Lee et al. evaluated sunlight redirecting panels designed to improve light penetration across the full height of cultivation racks, reporting an increase in photosynthetic activity of up to about 10 to 12% [15]. A second study proposed a façade-mounted, multi-layer reflector system combined with infrared cutting glazing to achieve a more uniform light distribution for vertically grown leafy crops. Under sunny day conditions, this configuration reduced electricity consumption by 33.8% compared with a reference vertical farm relying entirely on artificial lighting [16].

For these reasons, it can be more promising to investigate daylighting approaches that deliver natural light directly to each growing zone, rather than relying on limited façade or roof penetration. One strategy that is attracting increasing research interest is the use of optical fibers to capture sunlight outdoors and transport it to individual racks within a vertical farm. For instance, Vu et al. assessed an experimental system combining a solar concentrator with optical fibers and reported energy reductions of about 54% on a sunny day and 39% on a cloudy day [17]. Building on the same concept, Yalcin et al. paired fiber-based daylight transport with a light distribution system that uses fluorescent coatings to spread light more evenly across the cultivation area and to shift the spectrum toward wavelengths more suitable for plant growth, using specialized reflective materials inside the farm [18]. Using this configuration, Kaya et al. conducted a sensitivity analysis across different lighting strategies and found energy savings in the range of 22-29% compared with a reference vertical farm relying only on LED lighting [19].

The main limitations of these fiber-based concepts are the high cost of optical fibers and the additional complexity of the sun tracking systems required by solar concentrators. For this reason, the present study focuses on simpler daylighting solutions based on light pipes, as a practical pathway to reduce lighting electricity demand and improve techno-economic viability. Among the many daylight harvesting options, cost and maintainability are often decisive. Li et al. compared several active daylight harvesting systems and showed that tubular daylighting devices can be more cost-effective than optical fiber solutions in terms of delivered light per unit cost (lumens per dollar), supporting the adoption of simpler daylight transport technologies when economics are critical [20]. Furthermore, lessons from greenhouse light management can help ensure that replacing part of electric lighting with daylight does not introduce new thermal or spectral penalties [21]. Reviews on greenhouse claddings and smart covers show that solar radiation can be tuned using ultraviolet or near infrared blocking films to manage both heat gains and spectral quality [22]. For example, near infrared filtering can reduce the incoming heat load, although depending on the material, it may also reduce photosynthetically active radiation, so this trade-off must be carefully assessed [23].

In this context, a clear gap emerges in the daylighting literature for vertical farming. While several hybrid concepts have been explored, often relying on facade penetration, long optical paths, or comparatively complex collection and transport solutions, there is still limited evidence from year-round assessments that combine crop-responsive modelling with energy and economic metrics to quantify the real value of simple daylight transport in multi-tier vertical farms. To address this gap, this study investigates a roof-integrated daylighting configuration based on simple, straight tubular light pipes designed to deliver daylight directly to the upper cultivation tier. Compared with conventional tubular daylighting devices, the proposed LP concept is enhanced with a low-cost light-deflection subsystem consisting of an aluminum-coated mirror housed within a rotating dome, thereby improving the setting proposed by [24]. This arrangement is intended to maximize the fraction of direct solar radiation redirected toward the target growing zone. To promote uniform illumination across the cultivation area and reduce localized hotspots, a pyramidal prismatic diffuser is installed at the LP outlet.

High-efficiency tubular daylighting solutions developed for buildings, such as the system proposed by [20], combining a parabolic concentrator with a long duct and multiple apertures for office illumination, are not readily transferable to VF applications. Crops typically require the most spatially uniform photon flux possible to sustain predictable growth, and multi-aperture delivery along an extended duct can introduce strong gradients and non-uniformity at canopy level. The concept proposed here instead aims to increase optical efficiency while maintaining low system complexity, targeting performance improvements beyond optical-fiber approaches without resorting to expensive fibers or highly complex concentrator-based collection and distribution architectures.

The approach is evaluated on a container-scale vertical farm, 3 m in height, with three tiers per rack, as a meaningful benchmark for roof-integrated daylighting in compact urban systems. In this configuration, the roof area remains significant relative to the cultivated area and daylight can be delivered to the upper growing tier through short optical paths, limiting transport losses and system complexity. Daylighting is therefore restricted to the tiers directly served by the light pipes, indicating that extension of the concept to taller large-scale facilities would require redesign. Limiting daylight collection to the roof surface further reflects an urban deployment scenario, where surrounding buildings and crowding can substantially reduce daylight availability on vertical facades. Dubai was selected as a high-solar-resource case study that enables the daylighting concept to be tested under favorable daylight-availability conditions while also making the associated cooling penalty clearly observable. To enhance light pipe performance while retaining low system complexity, a simple tracking collector, a tilting aluminum-coated mirror inside a rotating dome, is simulated via ray tracing to increase daylight flux reaching the crop and improve spatial distribution at the target plane. The resulting daylight input is coupled with multiple artificial light supplementation strategies in year-round simulations using the validated dynamic AGRI-Energy model by [25], which predicts how time-varying light intensity affects plant growth and evapotranspiration, enabling consistent evaluation of biomass formation and the associated climatic loads [25]. Scenarios are compared in terms of crop yield, light use efficiency (LUE), and specific electric energy consumption (SEEC), and from an economic perspective through light cost (LC) and payback time (PBT). On this basis, the analysis captures the central techno-economic trade-off of hybrid daylight-electric lighting systems, namely the balance between lighting savings, additional thermal loads, yield preservation, and investment cost. Hence, the main scientific contributions of this work are summarized as follows.

- Development and assessment of a low-complexity daylighting concept for vertical farming, based on straight tubular light pipes integrated at roof level and optimized for direct roof-to-canopy delivery in multi-tier container-scale systems.
- Definition of an integrated optical–agronomic–energy evaluation framework, combining ray-tracing simulations with the AGRI-Energy model to consistently quantify daylighting impacts on key performance indicators.
- Comprehensive year-round techno-economic assessment of a roof-mounted light pipe system for small-scale multi-tier vertical farms, coupling daylight transport with crop-responsive dynamic modelling and providing quantitative evidence on light cost, payback time, and scalability limits in dense urban contexts.

## 2. Methodology

This section presents the assumptions and methods used to assess the techno-economic viability of integrating a light pipe system to reduce energy consumption in the container farm.

### 2.1. Vertical Farm AGRI-Energy Model

The results are based on the AGRI-Energy model by [25], shown in Figure *1*, which links indoor setpoints to both subsystem energy demand and crop growth within the growing chamber. Model inputs include outdoor conditions (temperature and solar irradiation), growing chamber setpoints (air temperature, relative humidity, $CO_2$ concentration, and photosynthetic photon flux density (PPFD)), and crop parameters. Model outputs are the power and energy demand of the heating, ventilation, and air conditioning system (HVAC), the water used by the VF, and the resulting biomass harvested. This integrated framework enables a detailed evaluation of how environmental control strategies interact with, and ultimately shape, overall vertical farm performance.

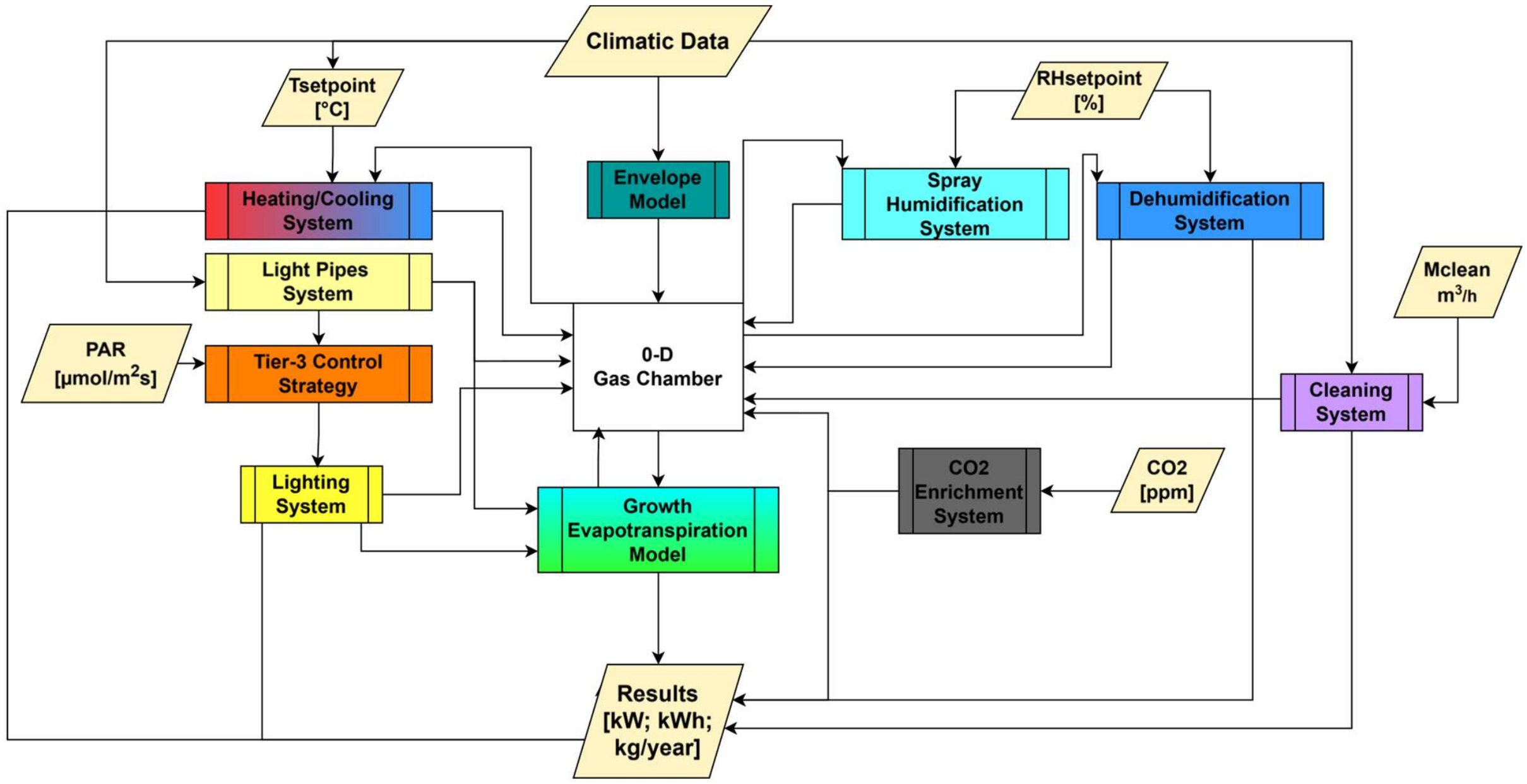

Figure 1: Flowchart of the employed model [25].

The chamber energy balance was extended, as shown in Eq. 1, to account for the LPs system by introducing additional contributions, the solar radiative gain transmitted through the LPs, $Q_{LP,sol}$, and the buoyancy-driven convective heat transfer within the LPs, $Q_{LP,conv}$, terms. When indoor air temperature exceeds the outdoor temperature, natural convection can develop inside the pipe and increase heat exchange with the external environment.

All remaining terms follow the validated AGRI-Energy formulation in [25]. $Q_{LED}$ represents the radiant power emitted by the LED system, while $Q_{plant}$ is the fraction intercepted by the crop and converted through photosynthesis. Sensible loads handled by the heating and cooling system and by the air handling unit are denoted by $Q_{h/c}$ and $Q_{AHU}$, respectively. Latent loads associated with moisture exchange are represented by $Q_{eva}$ for crop evapotranspiration and $Q_{hum}$ for humidification. Heat transfer through the chamber envelope is captured by $Q_{env}$. Details of the LP terms are provided in the next section, whereas the estimation of the remaining power terms is described in [25].

$$d_{air} c_{p_{air}} V_{room} \frac{\partial T}{\partial t} = Q_{env} + Q_{LED} + Q_{LP,sol} - Q_{LP,conv} - Q_{plant} - Q_{eva} + Q_{h/c} - Q_{AHU} - Q_{hum} \quad (1)$$

Lettuce was selected as the reference crop due to its widespread suitability for vertical farming. Biomass production at each timestep was calculated using the validated growth model reported in [25], which relates dry and fresh matter production to air temperature, $CO_2$ concentration and PPFD (Eqs. 2-3), through their influence on leaf area index (LAI) and light use efficiency (LUE). Baseline LUE values were taken from the experimental dataset of Carotti et al. [26] and then adjusted for different $CO_2$ levels using the data reported in [27], as illustrated in Figure *2*. Crop evapotranspiration was estimated following the same approach as in [25]. Since no modifications were introduced in this sub-model, further details are not repeated here.

$$\frac{\partial DM}{\partial t} = PPFD \cdot (1 - e^{-k \cdot LAI}) \cdot LUE_{dm} \cdot A_{crop} \quad (2)$$

$$\frac{\partial FM}{\partial t} = PPFD \cdot (1 - e^{-k \cdot LAI}) \cdot LUE_{fm} \cdot A_{crop} \quad (3)$$

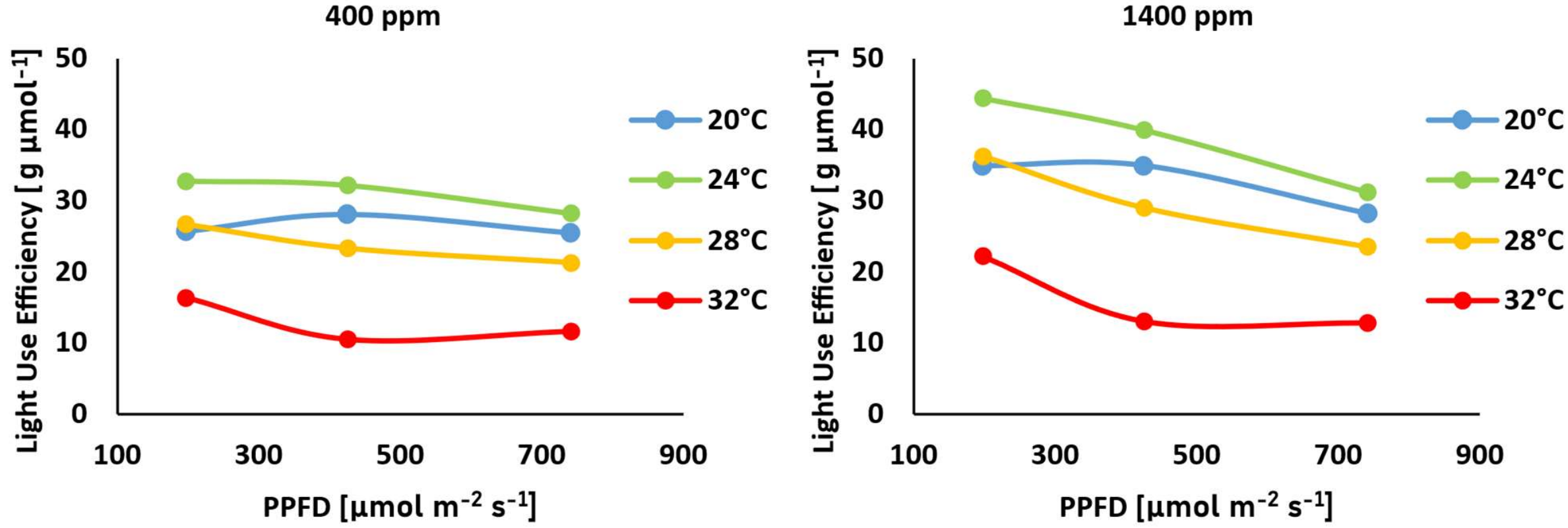

Figure 2: Experimental Light use efficiency of lettuce under different growing conditions.

### 2.2. Proposed Vertical Farm Configuration

The vertical farm considered in this study is a container farm with dimensions 7x7x3 m, suitable for urban deployment. The envelope is insulated with polyurethane (PUR) sandwich panels (U-value: 0.17 – 0.18 W $m^{-2}$ $K^{-1}$) to reduce transmission losses. The growing chamber was equipped with five-rack units, each with three tiers. The total cultivated area, computed as the sum over all tiers, is 90 $m^2$ as described in [25]. Lettuce crops were grown at a density of 25 plants $m^{-2}$ and harvested at an average fresh mass of 250 g per plant [26]. Indoor setpoints were selected based on the sensitivity analysis by [7], adopting the configuration that minimizes the levelized cost of lettuce (LCoL). Since relative humidity showed a limited influence on crop growth, RH setpoints were chosen to reduce condensation and microbial proliferation risk. Ultimately, the VF was assumed to be located in Dubai to take advantage of the high availability of natural light. In arid climates such as the UAE (United Arab Emirates), cultivation technologies that minimize freshwater demand are particularly attractive for enhancing food security under water-scarce conditions. In the analyzed VF, water-use efficiency is high because condensate recovered at the HVAC cooling coils is recirculated, so that most of the net water requirement is effectively embodied in the harvested biomass rather than lost to the environment. The climatic data used in the current study were taken from [28] and reported in detail in the Appendix. All design parameters and setpoints are summarized in Table *1*.

Table 1: VF design parameters

| Variable | Value | Unit |
| --- | --- | --- |
| **Container Farm Area** | 49 | $m^2$ |
| **Cultivation Area per Tier** | 30 | $m^2$ |
| **Overall Cultivation Area** | 90 | $m^2$ |
| **Crop Density** | 25 | plants $m^{-2}$ |
| **Envelope Thermal Transmittance** | 0.17 – 0.18 | W $m^{-2}$ $K^{-1}$ |
| **Indoor Air Temperature Setpoint** | 24 | °C |
| **$CO_2$ Level Setpoint** | 1400 | ppm |
| **PPFD Setpoint** | 250 | μmol $m^{-2}$ $s^{-1}$ |
| **Relative Humidity Setpoint** | 75% (light) / 85% (dark) | - |

This baseline configuration is used to quantify the impact of the light pipe system on the chamber energy balance and overall VF performance. As mentioned in the Introduction, daylight delivery to the lower tiers is not feasible. Hence, tiers 1 and 2 are equipped with LED lighting providing a constant PPFD of 250 μmol $m^{-2}$ $s^{-1}$ over a 16 h photoperiod, corresponding to a daily light integral (DLI) of 14.4 mol $m^{-2}$ $day^{-1}$. Conversely, daylight is supplied to tier 3 through roof-mounted LPs. Assuming one growing position per 20 x 20 $cm^2$ area as in [26], tier 3 includes 750 growing positions. Accordingly, one LP per position is assumed (750 LPs). To maximize the direct radiation delivered to each growing zone, each LP includes an internal aluminum-coated mirror. The mirror was introduced to reduce internal reflection losses within the light pipe, particularly when solar altitude is low, and its adjustable tilt allows the redirection effect to be adapted to changing solar position. Mirror tilt is adjusted according to solar altitude by a micro-servo motor mounted externally on the dome. Dome rotation follows solar azimuth and is achieved through a geared transmission. Each dome is fixed to a toothed wheel that is mechanically coupled to adjacent domes through intermediate idler gears, forming a continuous gear train along the row. This configuration allows a single stepper motor, located at one end of the row, to actuate all LPs simultaneously. Shared actuation is justified because all domes within a row require the same azimuthal orientation. The shared actuation strategy reduces the number of motors required and provides a simpler control architecture for the row-level tracking system. Because actuation is intermittent and involves only small angular adjustments, the associated auxiliary electricity demand is expected to be negligible compared with the VF energy use. From a design perspective, the high number of LP units makes modular assembly, roof sealing, component accessibility, and weather-resistant encapsulation key aspects for future prototype optimization. These elements are also relevant to reduce the investment cost, which represents the main barrier identified in the economic assessment. To limit the external footprint of the active collection system, the LP diameter is set to 150 mm, while the pipe length is assumed equal to 1 m as a conservative choice. At the outlet, a pyramidal-shaped prismatic diffuser spreads the rays and improves illumination uniformity over the target area. Diffuser dimensions were selected to maximize uniformity at the crop level. The canopy distance is set to 0.5 m, consistent with typical artificial lighting layouts. Finally, external components are assumed to be protected by an encapsulation system to mitigate weather-related degradation. The overall container farm configuration and LP details are shown in Figure *3*, while Table *2* summarizes the main assumptions for the LP system.

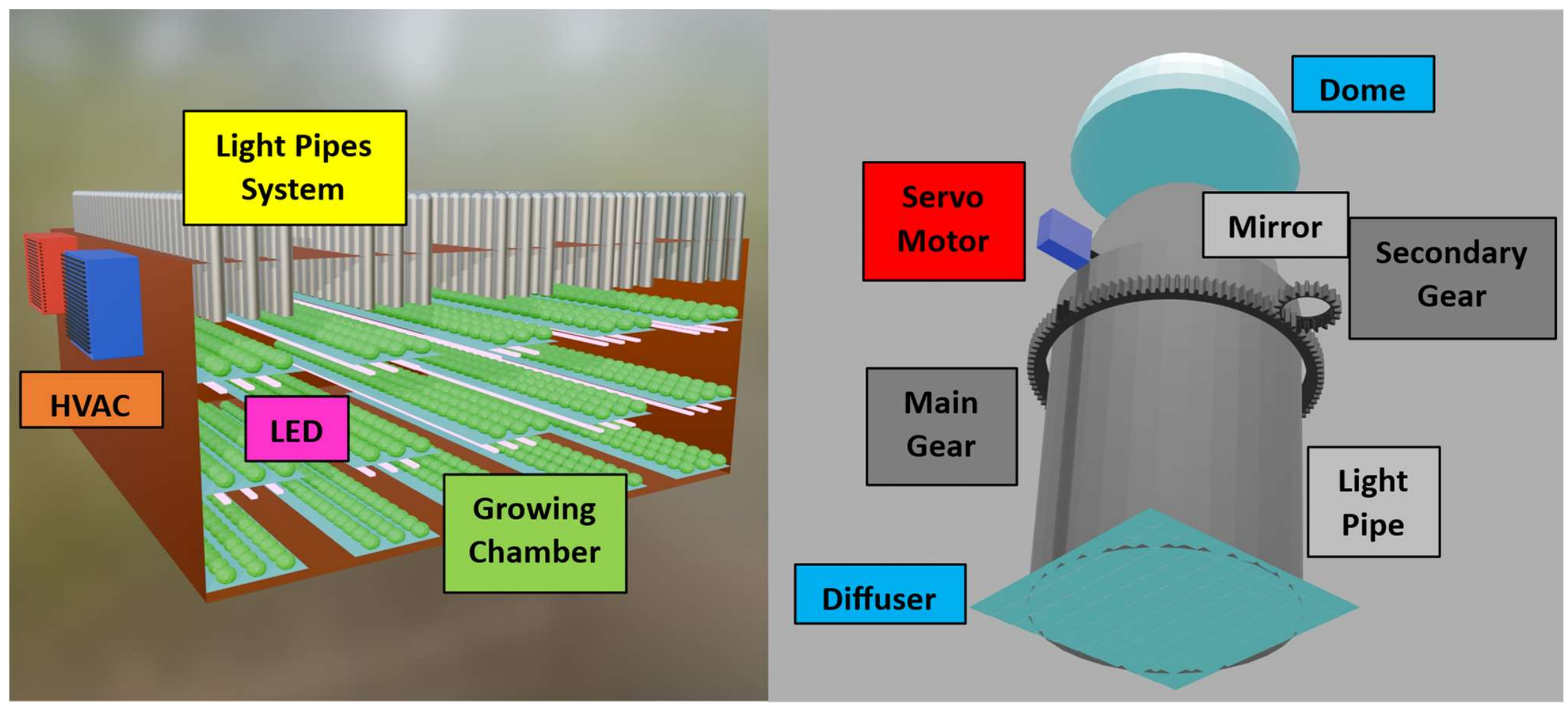

Figure 3: 3D model of the proposed daylighting strategy (left) and the LP layout with the sun-tracking system (right)

Table 2: Key assumptions of the proposed daylighting strategy.

| Variable | Value | Unit | Reference |
|---|---|---|---|
| **LP Diameter** | 150.0 | mm | [-] |
| **LP Length** | 1,000.0 | mm | [-] |
| **Dome Thickness** | 2.0 | mm | [-] |
| **Dome Transmittance** | 0.91 | [-] | [29] |
| **LP Internal Reflectance** | 0.90 | [-] | [30] |
| **Mirror Reflectance** | 0.90 | [-] | [31] |
| **Diffuser Thickness** | 2.0 | mm | [-] |
| **Diffuser Prism Pitch** | 2.0 | mm | [-] |
| **Diffuser Pyramid Height** | 4.0 | mm | [-] |
| **Diffuser Pyramid Apex Angle** | 152 | ° | [-] |
| **Total LPs Area** | 13.25 | $m^2$ | [-] |

### 2.3. Light Pipe System

To quantify the direct and diffuse radiation delivered to the growing zone located beneath each LP, the full optical system, including dome, pipe, mirror, and diffuser, was modelled using the ray-tracing software Tonatiuh. Simulations were performed for different solar positions to evaluate how solar altitude and azimuth affect the crop-level LP optical efficiency of the system and the resulting light distribution at crop level. The modelling of refractive components follows the approach reported in [32] and an example of the optical model is shown in Figure *4*. For each LP, the solar gain is calculated as the sum of a direct and diffuse component, as expressed in Eq. 4. Direct and diffuse radiation were treated separately using the DNI and DHI components of the climatic dataset [28]. This distinction was retained throughout the optical and energy calculations by assigning separate Tonatiuh-derived optical efficiencies to the direct and diffuse components, since they interact differently with the mirror, pipe walls, and diffuser.

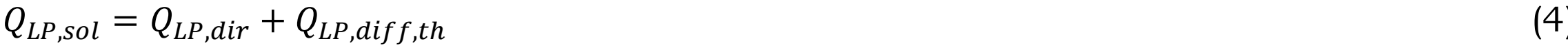

$$Q_{LP,sol} = Q_{LP,dir} + Q_{LP,diff,th} \tag{4}$$

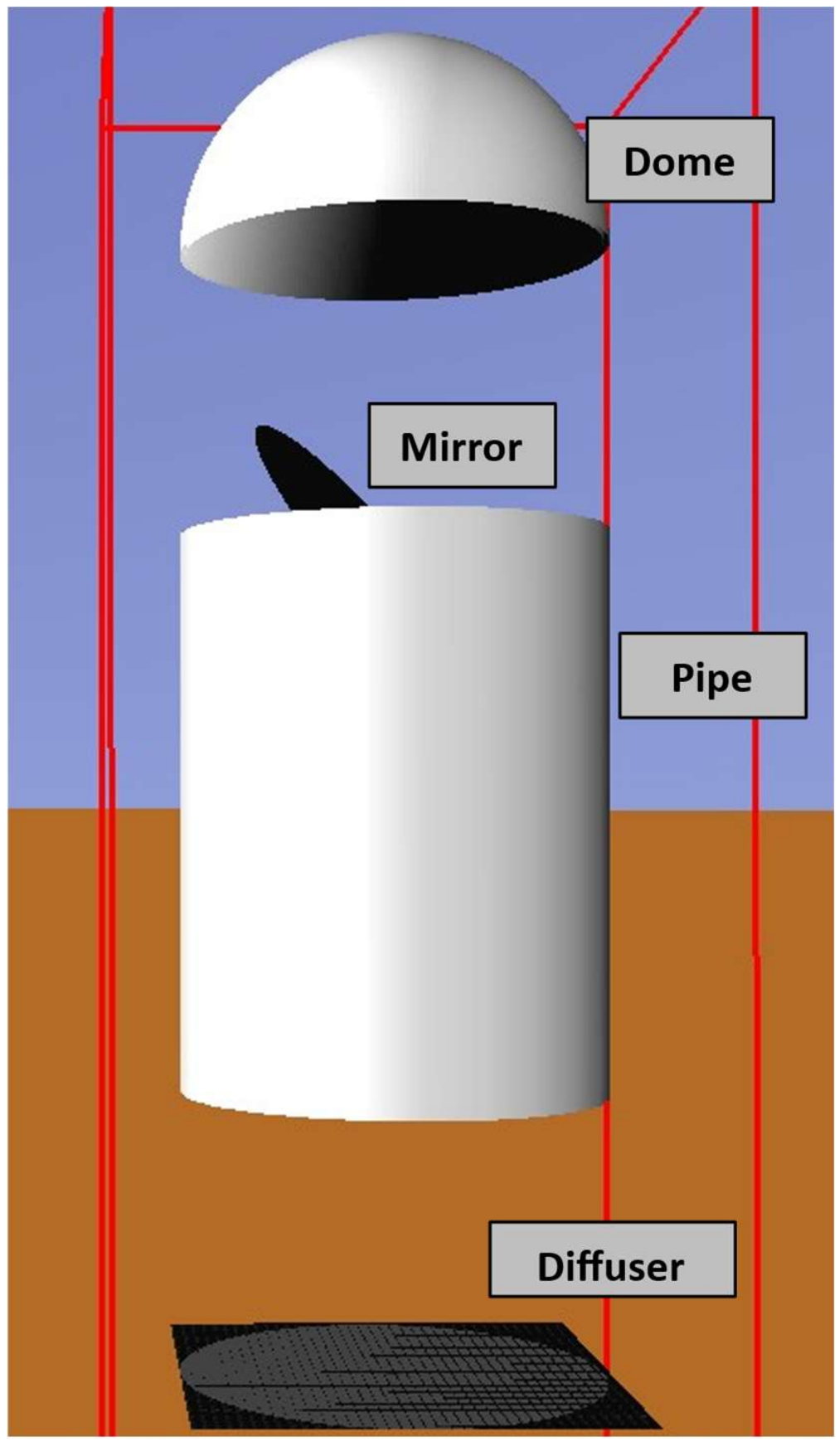

Figure 4: Light-pipe optical setup modelled on Tonatiuh.

The direct component has been estimated as expressed in Eq. 5. Where $I_{dir}$ is the direct normal irradiance (DNI), θ is the incidence angle, and $\eta_{dir}$ is the Tonatiuh-derived optical efficiency defined as the ratio between the power reaching the target growing zone and incident beam power on the horizontal collector aperture $A_{LP}$. Solar altitude and azimuth were computed for Dubai, and the corresponding incidence angles are reported in the Appendix.

$$Q_{LP,dir} = I_{dir} \cdot \cos\theta \cdot \eta_{dir} \cdot A_{LP} \tag{5}$$

Diffuse horizontal irradiance (DHI) is assumed to be isotropically distributed over the sky hemisphere. The dome is discretized into ten-degree-wide angular bands, and Tonatiuh is used to compute the transmitted power for each band as a function of mirror tilt. Since the mirror occludes the rear half-dome, the diffuse contribution is halved for bands between 10° and 90°. Here, $\eta_{diff,th,i}$ and $\eta_{diff,crop,i}$ denote the Tonatiuh-derived, band-wise diffuse optical efficiencies for the $i$-th angular sector of the dome, computed respectively for the total transmitted power contributing to the chamber energy balance and for the fraction reaching the target growing zone used in the crop-growth submodel. The diffuse term used in the energy balance is defined as Eqs. 6-7. Because mirror tilt is optimized for direct radiation, only a fraction of diffuse light reaches the target area. Therefore, a second diffuse efficiency is introduced for the crop growth sub-model and employed in Eq. 8.

$$Q_{LP,diff,th} = \sum_0^{\pi/2} \frac{I_{diff}}{2} \cdot f_{dome,i} \cdot \eta_{diff,th,i} \cdot A_{LP} \tag{6}$$

$$f_{dome,i} = (\sin i)^2 - (\sin(i - \frac{\pi}{18}))^2 \quad (7)$$

$$Q_{LP,diff,crop} = \sum_{10}^{90} \frac{I_{diff}}{2} \cdot f_{dome,i} \cdot \eta_{diff,crop,i} \cdot A_{LP} \quad (8)$$

In this study, the reported crop-level LP optical efficiency is defined as the fraction of radiation incident on the collector aperture that reaches the target growing zone at crop level and is therefore used as daylight input to the crop-growth model. The resulting PPFD contribution from the LP system is then computed as expressed in Eq. 9. The energy is converted to µmol $m^{-2}$ $s^{-1}$ by dividing by the crop area and using a solar-spectrum conversion factor of 2.247 µmol $J^{-1}$ [33].

$$PPFD_{LP} = (Q_{LP,dir} + Q_{LP,diff,crop}) \cdot \frac{2.247}{A_{crop}} \quad (9)$$

The LPs also affect roof transmission losses through buoyancy-driven convection within the pipes. When $T_{in} <= T_{ext}$, natural convection is suppressed. When $T_{in} > T_{ext}$, the convective heat transfer is computed as described in Eqs. 10-13, where L is the pipe length, and $A_{LP}$ is the heat-transfer area associated with each LP.

$$Ra = \frac{g \cdot \beta \cdot \Delta T \cdot L^3}{\frac{\nu^2}{Pr}} \quad (10)$$

$$Nu = 0.15 Ra^{0.33} \quad (11)$$

$$U_{LP} = \frac{Nu \cdot k_{air}}{L} \quad (12)$$

$$Q_{LP,conv} = U_{LP} \cdot A_{LP} \cdot \Delta T \quad (13)$$

### 2.4. Sensitivity Analysis

To investigate in detail the impact of the LP system, a set of scenarios was analyzed. The "Bench" case represents the benchmark configuration with no LPs, where all tiers are illuminated exclusively by LEDs. The "LP_NL" case evaluates daylighting only, with tier 3 illuminated solely by the LP system, while tiers 1 and 2 rely on LEDs.

A first level of LP-LED integration is explored in the "LP_Min" scenario, where LEDs on tier 3 operate with an on/off logic during a 16 h photoperiod (4 a.m. to 8 p.m.). In this case, LEDs are activated when the PPFD delivered by the LPs falls below a minimum threshold of 100 µmol $m^{-2}$ $s^{-1}$, and two nominal LED PPFD setpoints were tested when ON, namely 200 and 250 µmol $m^{-2}$ $s^{-1}$.

A more advanced integration strategy is considered in the "LP_Dim" scenario, where a dedicated driver enables PWM regulation of tier-3 LEDs to complement daylighting and achieve a target PPFD of 250 µmol $m^{-2}$ $s^{-1}$ over the same photoperiod [34][24]. PWM dimming was selected because it preserves the LED spectral distribution, whereas current-controlled dimming may increase electrical efficiency at part load, but can alter the emitted spectrum [35]. The efficiency penalty associated with PWM operations is included by assuming a driver efficiency of 95% [36] and applying the part-load efficiency dependence reported in [37] (see Figure 5), with a minimum dimming level of 30%.

Since a fraction of the incoming daylight may not be beneficial for crop growth, two additional variants of "LP_Dim" were tested. In "LP_Dim_IR", a UV-IR filter is introduced in the LP optical path. Two visible range optical transmissions were considered (98% and 90% [17]), yielding overall efficiencies of 44.7% and 41.0%, respectively. Under this assumption, the transmitted radiation is entirely within the photosynthetically active radiation (PAR) range (400-700 nm), and therefore conversion factor of 4.56 µmol $J^{-1}$ [33] is adopted. In "LP_Dim_EC", daylighting is regulated through a polymer-dispersed liquid crystal film actuator, whose optical transmittance is controlled by the applied voltage, following the approach described in [38]. The control objective is to limit excessive incident PPFD at the crop level, with an upper bound set to 400 µmol $m^{-2}$ $s^{-1}$ to avoid reduced LUE and potential stress. The voltage-transmittance relationship is described by Eq. 14.

To assess the robustness of the results, an additional sensitivity analysis was performed on the selected LP_Dim scenario by reducing the effective crop-level LP optical efficiency by 10% and 20%. These reductions were introduced as a simplified representation of combined non-ideal effects in the optical subsystem, such as tracking inaccuracies, installation deviations, dust accumulation, aging, and departures from nominal optical properties. The corresponding impact on the main energetic indicators was then evaluated.

Finally, the "GH" scenario replaces the roof and part of the external walls with greenhouse-like glazing, providing a simplified daylighting reference for comparison with the LP approach. In this case, the glazing is characterized by a thermal transmittance of 3.75 W $m^{-2}$ $K^{-1}$ and an average optical transmittance over the solar spectrum of 0.82 [39]. Also in this case, only tier 3 is affected by daylighting, to ensure consistency across scenarios. All scenarios were simulated under three values of LED photosynthetic photon efficiency (PPE) to assess how the relative benefit of daylighting changes with lighting technology. A value of 3.0 µmol $J_{el}^{-1}$ represents state-of-the-art LEDs, while 2.5 and 2.0 µmol $J_{el}^{-1}$ reflect lower efficiency solutions commonly adopted in commercial container farms. Table *3* summarizes the scenario definitions and tier-3 control logic.

$$\tau_{EC} = \frac{0.1331 \cdot v^2 - 0.5184 \cdot v + 8.4437}{0.1811 \cdot v^2 - 0.8825 \cdot v + 15.5613} \tag{14}$$

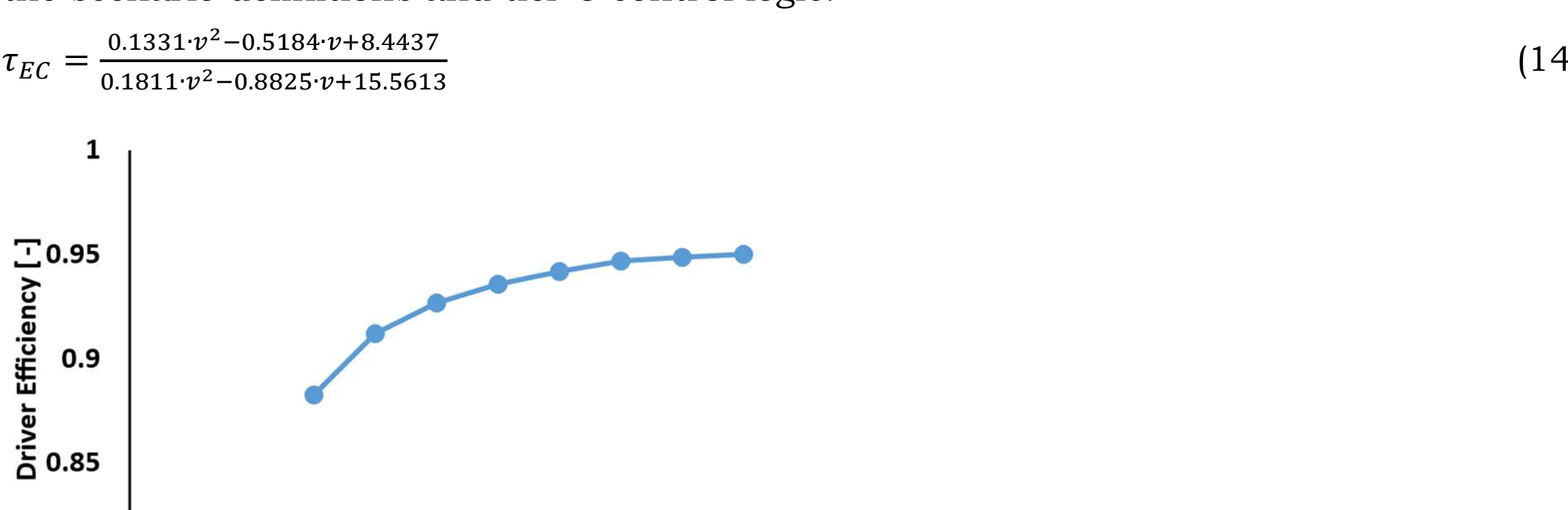

Figure 5: LED driver efficiency at part load.

Table 3: Scenarios Investigated

| **Scenario** | **Tier-3 lighting/control strategy** | **Target PPFD [µmol $m^{-2}$ $s^{-1}$]** |
|---|---|---|
| **Bench** | LED Only | 250 |
| **LP_NL** | LP Daylighting Only | N/A |
| **LP_Min_200** | On/off LEDs when LP PPFD < 100 | 200 |
| **LP_Min_250** | On/off LEDs when LP PPFD < 100 | 250 |
| **LP_Dim** | PWM dimming to meet setpoint | 250 |
| **LP_Dim_IR_98** | PWM dimming + UV–IR filter ($\tau_{vis}$ = 0.98) | 250 |
| **LP_Dim_IR_90** | PWM dimming + UV–IR filter ($\tau_{vis}$ = 0.90) | 250 |
| **LP_Dim_EC** | PWM dimming + variable transmittance (PPFD cap: 400) | 250 |
| **GH** | Common Glazing | N/A |

### 2.5. Economic Assumptions

To evaluate the economic viability of the proposed system, the most energy-efficient scenarios were compared against the benchmark from an economic standpoint. Cost assumptions for LP implementation, including LP components, filters, and LEDs, are reported in Table A1 (Appendix). In particular, the unit cost of a light pipe, including the mirror and the rotating mechanism, was assumed to be 300 \$ $unit^{-1}$ [40], [41]. This value is slightly lower than the 333 – 666 \$ $unit^{-1}$ range reported in [20], because the proposed design does not require high-cost internal materials due to the mirror-based redirection, and the pipe length is approximately one sixth of that considered in the reference.

Payback time (PBT) was computed for each selected scenario by comparing the additional investment to illuminate tier 3 using LP-based solutions against the reference case relying on LED lighting only. The incremental CAPEX also accounts for difference in HVAC sizing associated with the additional thermal load introduced by solar radiation transmitted through the LP system [42]. Since LP-based daylighting can affect VF productivity, revenues were adjusted using a constant lettuce selling price obtained from [2]. A sensitivity analysis on electricity price, carbon taxation, and LP unit cost was then carried out to quantify the dependence of cost-competitiveness on the main uncertain economic drivers. Electricity price and carbon taxation were considered because they directly affect the value of the energy savings, whereas LP unit cost was included to account for the main uncertainty associated with the additional CAPEX of the proposed system. The role of initial investment was further explored by identifying the break-even LP unit cost required to achieve an acceptable PBT. Finally, under the same available daylight level (100,000 lux), the performance of the proposed LP system was computed and compared with the optical fiber solution proposed by [17], including a comparison in terms of light cost. The economic assessment was carried out as a first-order techno-economic screening of the LP-based configurations against the LED-only benchmark. The adopted CAPEX values were considered as installed costs, including both equipment and installation. Since the proposed LP system relies on relatively simple mechanical and optical components, auxiliary electricity use for dome actuation, maintenance requirements, and performance degradation were assumed to have a secondary impact compared with LP CAPEX and electricity-cost savings.

## 3. Results

### 3.1. Light Pipes System

In the proposed configuration, each LP includes a dome-mounted, aluminum-coated mirror designed to steer the incoming beam towards the target growing zone, rather than distributing daylight over a broad area as in conventional daylighting applications. Accordingly, ray-tracing simulations performed in Tonatiuh indicate that the optical performance of the LP system depends strongly on solar altitude. Figure 6 reports the direct and diffuse crop-level optical efficiencies of the LP system, accounting for the multi-LP roof layout and the potential overlap of the illuminated footprints. The mirror is effective for the direct beam, increasing $\eta_{dir}$ under DNI by steering collimated rays toward the growing zone. In contrast, diffuse radiation is weakly coupled to the mirror geometry, and it is predominantly attenuated or redirected away from the target by mirror occlusion and multiple interactions with the internal surfaces of the pipes. As an upper bound for a single LP, the maximum theoretical efficiency for DNI is 73%, limited by optical losses through the dome, mirror, and diffuser. Conversely, the maximum efficiency under DHI is 37% since the mirror inherently occludes approximately half of the dome hemisphere.

The trend of $\eta_{dir}$ shows a pronounced reduction for solar altitude between 40° and 60°, which can be traced to the coupled interaction between mirror interception and diffuser angular response. At low solar altitude, the mirror exhibits a larger apparent area relative to the LP horizontal aperture. Accordingly, the mirror interception ratio, defined as the power intercepted by the mirror divided by the incident power on the LP horizontal aperture, can exceed 100%. This does not indicate an optical efficiency greater than unity, but simply reflects the fact that, at low solar altitude, the apparent projected area of the mirror can be larger than the horizontal aperture area used as the reference in the ratio definition. As solar altitude increases, the apparent area of the mirrors decreases, and an increasing fraction of direct rays bypasses the mirror. These rays undergo multiple internal reflections within the pipe and reach the diffuser with non-ideal incidence angles. Since the diffuser geometry is designed for near vertical rays, a substantial portion of this radiation is refracted away from the target growing zone, leading to the observed drop in $\eta_{dir}$ at intermediate altitudes. At higher solar altitude, although the fraction of mirror intercepted energy keeps decreasing, the incident rays are closer to vertical and experience fewer internal reflections. Consequently, the diffuser redirects a large share of the transmitted radiation toward the crop and $\eta_{dir}$ recovers.

The $\eta_{diff}$ follows the opposite behavior, with maximum diffuse contribution transmitted into the growing chamber occurring at a solar altitude of about 50°. This condition corresponds to a mirror tilt of 70°, which is an intermediate position within the adopted tracking range. When the sun is low, the mirror tilt is small, and a large fraction of the diffuse radiation is intercepted by the rear side of the mirror and thus lost. When the sun is high, the mirror approaches a near-vertical orientation, and diffuse rays are more likely to undergo multiple reflections within the pipe, because the optical layout is primarily designed to redirect the direct beam. A further relevant outcome is the marked difference between $\eta_{diff,th}$ and $\eta_{diff,crop}$. Most diffuse radiation that enters the system is not delivered to the target growing zone and is instead spread by the diffuser toward surrounding areas. Since this redistribution among neighboring growing zones cannot be quantified reliably within the present modelling framework, a conservative assumption was adopted by neglecting this contribution in the crop growth sub-model. As a result, only 17 to 31%, depending on solar altitude, of the diffuse radiation entering the growing chamber is considered to effectively contribute to crop cultivation. The effect of this assumption on the final indicators is limited by the relatively small contribution of diffuse radiation to the LP energy input. Over the simulated year, the diffuse component accounted for approximately 8% of the total transmitted energy in summer and up to 24% in winter. Moreover, even under a non-conservative assumption in which all diffuse radiation transmitted into the chamber was considered useful for crop growth, the resulting variations in annual productivity and SEC would be only about 0.8% and 2.5%, respectively.

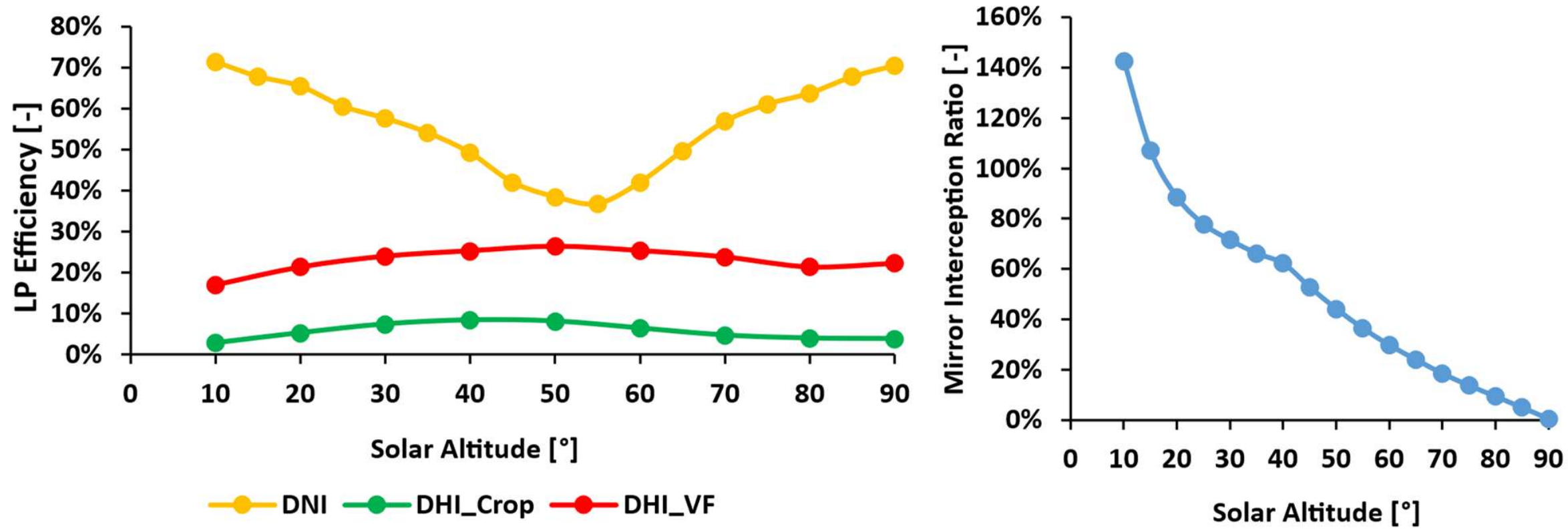

Figure 6: Crop-level optical efficiency of the LP system by radiation component (left) and mirror-intercepted area (right). The DHI terms refer to the diffuse radiation reaching the crop beneath the light pipe (DHI_Crop) and to the total diffuse radiation entering the farm (DHI_VF), respectively.

The system efficiency of LPs is not the only parameter affected by solar altitude. As shown in Figure *7*, solar altitude also influences the spatial light distribution over the target growing zone. At higher solar altitude, the mirror becomes less effective at redirecting the direct beam, leading to a more concentrated footprint and some local peaks. Nevertheless, the distribution remains more uniform than configurations without the internal mirror or the pyramidal-shaped diffuser. The present framework evaluates the footprint of a single LP and therefore does not explicitly capture the additional smoothing effects that may arise from overlap among neighboring LPs and from redistributed diffuse radiation.

Overall, the proposed system combines acceptable uniformity with high optical throughput. Under the same incident condition adopted in [17], namely an external DNI of 10,000 lux corresponding to 833 W m$^{-2}$, the LPs achieve a crop-level LP optical efficiency of about 75%, compared to 16-19% reported for the optical fiber system in [17], highlighting the advantage of a shorter optical conversion chain.

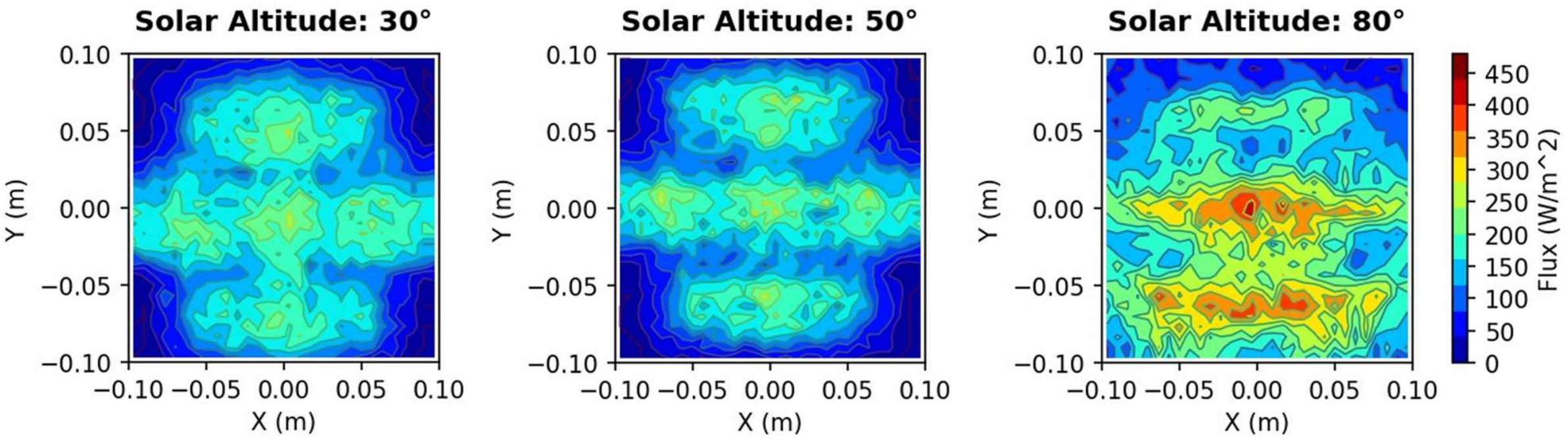

Figure 7: Distribution of natural light delivered by the LP system at the crop level for different solar altitudes.

The PPFD delivered to the crop strongly varies with the adopted daylighting strategy. As shown in Figure *8*, the Bench case maintains a constant PPFD of 250 µmol $m^{-2}$ $s^{-1}$ over the 16 h photoperiod. Without LED integration (LP_NL), PPFD is markedly lower during the morning and evening hours, while around noon the LP contribution can exceed the benchmark setpoint, reaching approximately +20% in winter and +110% in summer.

When LED supplementation is controlled through PWM-dimming (LP_Dim), the delivered PPFD closely follows the setpoint whenever daylight is absent or low, because the LEDs compensate for the deficit. Deviations from the target occur when daylight approaches the setpoint. Due to the minimum dimming level of the driver (30%), once daylight exceeds about 175 µmol $m^{-2}$ $s^{-1}$, the LEDs cannot be further reduced and are switched off, producing short mismatches relative to the setpoint. When daylight exceeds the target, LEDs remain off, and PPFD coincides with the daylight-only (LP_NL) profile.

Optical filtering is introduced to reduce solar gains and the associated cooling penalty. LP_Dim_IR_90 and LP_Dim_IR_98 attenuate the non-PAR UV-IR portions of the spectrum and reduce the transmitted radiant load. As a result, the PPFD profile retains the same temporal pattern as LP_Dim but is shifted downward due to additional optical losses introduced by the filter. The variable-transmittance scenario (LP_Dim_EC) further limits peak PPFD by enforcing an upper bound, by electronically reducing the transmittance of the lens. However, the lower average transmittance reduces PPFD over a larger fraction of the day, which leads to more frequent LED operation while still avoiding operation below the minimum dimming level, as evident in Figure 8 (c).

In the LP_Min configurations (LP_Min_250 and LP_Min_200), tier-3 LEDs operate with an on/off control based on a threshold of 100 µmol $m^{-2}$ $s^{-1}$. When daylight is absent, LEDs provide their nominal PPFD of 250 or 200 µmol $m^{-2}$ $s^{-1}$, respectively. When daylight is present but remains below the threshold, LEDs stay on, and the combined PPFD exceeds the setpoint, reaching up to about 349 µmol $m^{-2}$ $s^{-1}$. This behaviour explains the morning and evening peaks observed in the LP_Min profiles.

The GH configuration yields substantially higher PPFD (see Appendix) due to the larger glazed area. However, the maximum achievable daylight utilization is inherently limited to approximately 49%. This upper bound arises from the combined effects of glazing transmissivity (80%) and the fact that only 61% of the total surface area is occupied by cultivation tiers. This apparently easier solution resulted in agro-energetic considerations discussed in the following sections.

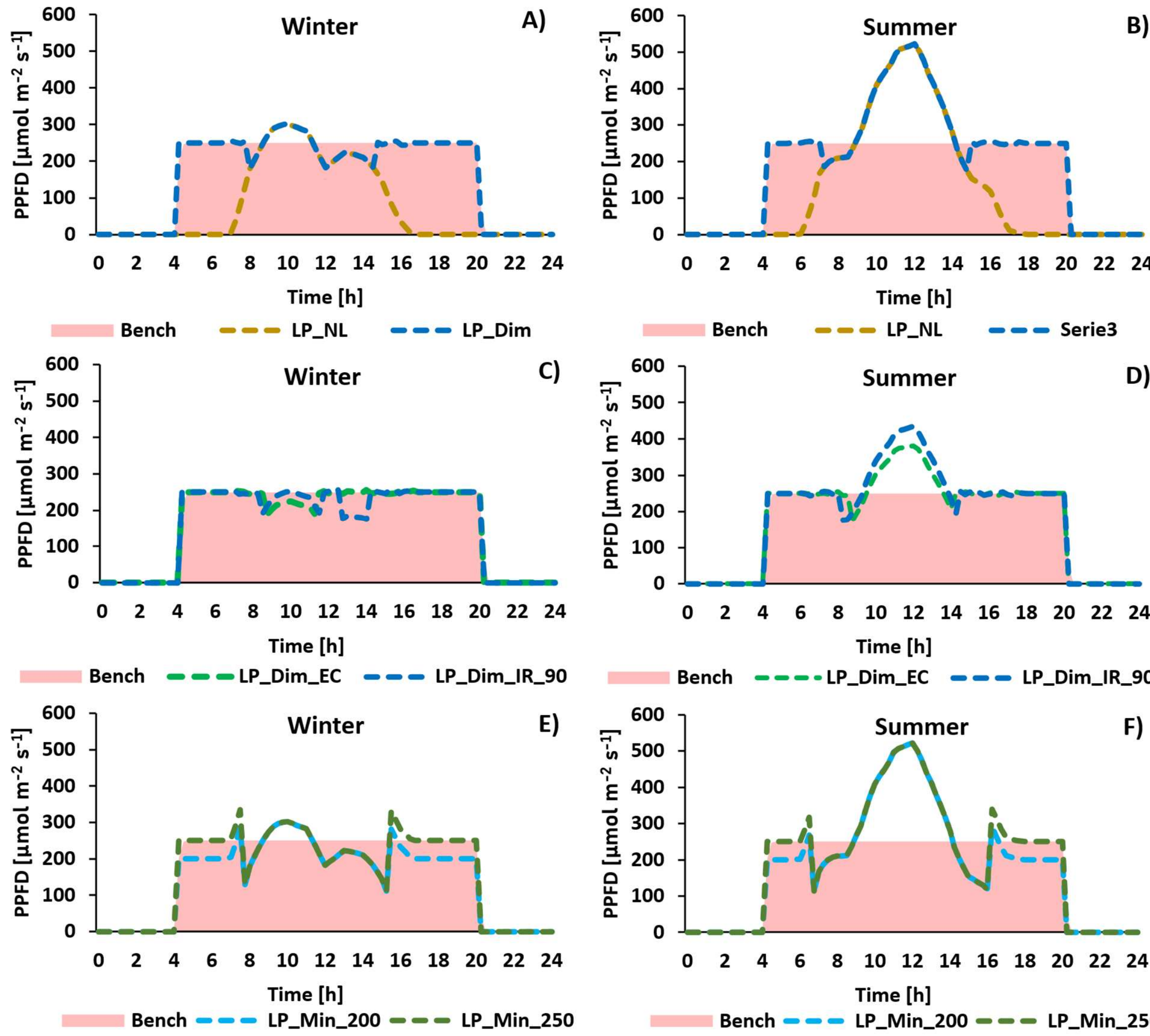

Figure 8: Crop-level PPFD under different LP daylighting strategies and seasonal conditions. Bench denotes the LED-only baseline. LP_NL is daylighting only on tier 3. LP_Min_200/250 are hybrid on/off strategies on tier 3 with nominal LED PPFD of 200 or 250 µmol $m^{-2}$ $s^{-1}$. LP_Dim denotes PWM dimming supplementation to meet the 250 µmol $m^{-2}$ $s^{-1}$ setpoint. LP_Dim_IR_90/98 indicate LP_Dim with UV–IR filtering, with visible transmittance 0.90 or 0.98. LP_Dim_EC denotes LP_Dim with variable transmittance control to limit peak PPFD.

### 3.2. LED Efficiency

LED PPE is a key determinant of container farm electricity demand, since artificial lighting accounts for the largest share of consumption. As reported in Table *4*, lighting represents approximately 77% of the total electricity use across all scenarios. Lower PPE not only increases lighting electricity demand but also raises cooling requirements due to higher waste-heat generation. Accordingly, reducing PPE by 16% and 33% increases the specific electricity energy consumption (SEEC) by 20% and 50%, respectively. Conversely, heating demand, dehumidification load, crop production, and water use efficiency (WUE) show negligible sensitivity to PPE in the analyzed cases, therefore, these metrics are not reported in Table *4*. In real-world applications, the economic advantage of low-cost LEDs depends on the electricity price, lower PPE reduces capital expenditures (CAPEX) but increases electricity consumption, which can dominate total costs over the system lifetime.

Table 4: Effect of PPE on the energy performance of the baseline VF (Bench)

| Variable | PPE 2.0 µmol $J^{-1}$ | PPE 2.5 µmol $J^{-1}$ | PPE 3 µmol $J^{-1}$ | Unit |
|---|---|---|---|---|
| **Lighting Nominal Power** | 11.3 | 9.0 | 7.5 | kW |
| **HVAC Nominal Power** | 11.1 | 9.2 | 8.3 | kW |
| **Cooling Energy** | 59.5 | 47.8 | 39.9 | MWh $year^{-1}$ |
| **Lighting Energy** | 66.0 | 52.8 | 44.0 | MWh $year^{-1}$ |
| **Electricity Requirement** | 85.1 | 68.0 | 56.7 | MWh $year^{-1}$ |
| **Crop Yield** | 9,221 | 9,221 | 9,221 | kg |
| **SEC** | 13.81 | 11.08 | 9.27 | kWh $kg^{-1}$ |
| **SEEC** | 9.22 | 7.38 | 6.15 | kWh $kg^{-1}$ |

### 3.3. Agronomic Standpoint

In this section, results refer to an LED lighting system with a PPE of 2.5 µmol $J^{-1}$. From an agronomic standpoint, daylighting has a negligible effect on the Water Use Efficiency because the HVAC system recovers most of the water removed during cooling and dehumidification. Here, the total lighting energy consumption is defined as the total lighting energy used for production, including both artificial and harvested daylight, per unit of mass of lettuce produced.

As shown in Figure *2*, at 24°C, crop growth is more efficient under a lower PPFD level. Accordingly, scenarios that limit excessive irradiance achieve lower total lighting energy consumption. The filtered configurations (LP_Dim_IR_98 and LP_Dim_IR_90) yield the lowest consumption, reaching 5.40 kWh $kg^{-1}$, by removing non-productive portions of solar radiation without affecting biomass production. The variable-transmittance strategy (LP_Dim_EC) reduces the harvested daylight by about 25% in winter due to the lower optical transmittance of the polymer-dispersed liquid crystal film. In summer, transmittance is further reduced by activating the filter, leading to a daylight reduction of up to about 40% to limit excessive, less efficiently utilized PPFD at crop level. This behaviour is reflected in Figure *9* by the increasing separation between the LP_Dim_EC and LP_NL PPFD profiles. Conversely, unfiltered configurations (LP_NL, LP_Min, and LP_Dim) exhibit higher total lighting energy consumption values, around 6.02 kWh $kg^{-1}$, because they do not attenuate either the non-photosynthetically active radiation (PAR) fraction of the solar spectrum or periods of high PPFD that reduce lettuce light use efficiency. The GH configuration admits substantially more daylight energy, but a large share is delivered at high PPFD and is therefore converted less efficiently into biomass, resulting in the highest consumption, approximately 16 kWh $kg^{-1}$, 182% higher than the Bench scenario (see Table *5*).

In energy terms, the LP system captures 11.1 MWh over the analysis period, corresponding to about 12% of the overall solar energy incident on the roof surface. This fraction decreases to 9% and 5% when optical filtering is applied, as part of the incoming radiation is intentionally rejected. Figure *10* shows the annual profile of captured daylight for the different strategies. Because the GH scenario admits daylight through the entire roof, it intercepts substantially more solar energy than LP-based solutions, whose effective collector area corresponds to about 27% of the roof. The remaining scenarios are not shown because they share the same harvested daylight profile and differ only in the LED supplementation and control strategy.

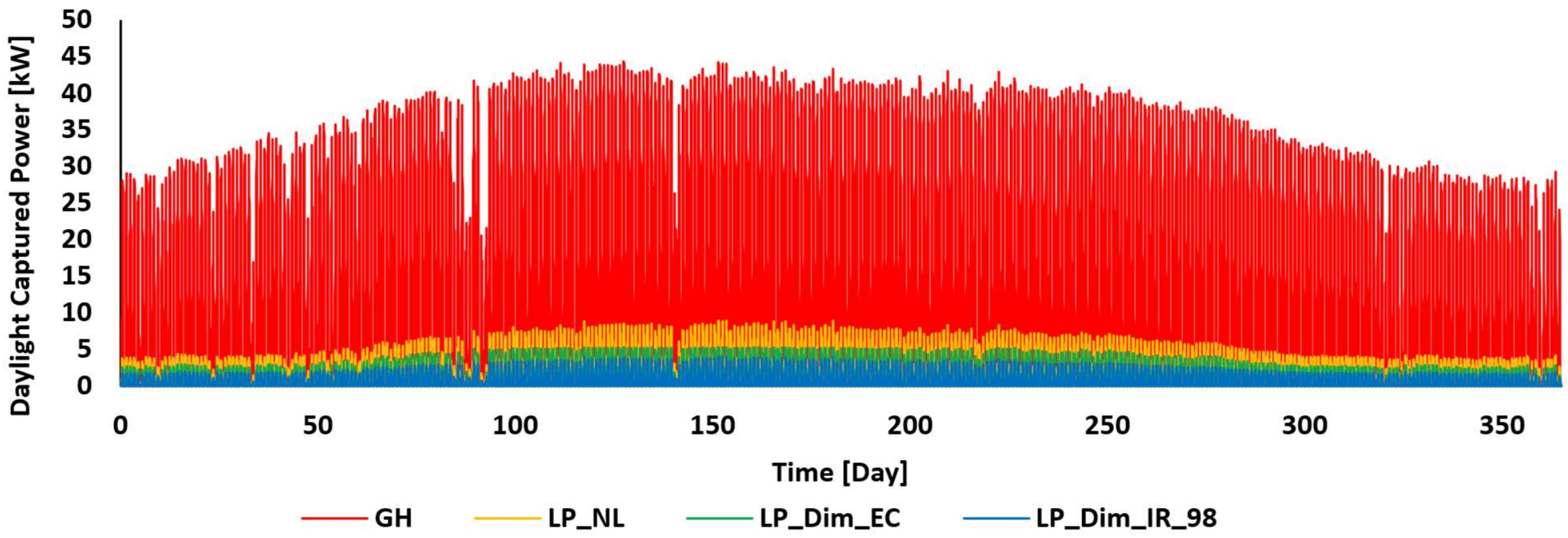

Figure 9: Comparison of captured daylight power across light-collecting systems. GH denotes the greenhouse-like glazing configuration. LP_NL denotes the light-pipe, daylight-only case on tier 3. LP_Dim_IR_98 denotes the light-pipe system with UV–IR filtering at visible transmittance 0.98. LP_Dim_EC denotes the light-pipe system with variable-transmittance control using an electrochromic actuator.

As a consequence, whereas artificial lighting provides a constant DLI of 14.4 mol $m^{-2}$ $day^{-1}$, daylighting produces a seasonally varying DLI over the simulated year. As shown in Figure *10*, the daylighting-only strategy (LP_NL) results in lower light exposure, with DLI reduction of 53% in winter and 8.5% in summer. Conversely, LPs integrated with LEDs through on/off or dimming control (LP_Min and LP_Dim) maintain a winter DLI comparable to the benchmark and increase summer DLI by about 18%. Introducing variable-transmittance control (LP_Dim_EC) reduces DLI by 7% relative to the unfiltered dimming case (LP_Dim) in both winter and summer. Conversely, the GH configuration reaches peak DLIs up to 56 mol $m^{-2}$ $day^{-1}$. Such high light exposure is unlikely to yield proportional productivity gains due to photosynthetic saturation and may adversely affect product quality, for instance, by increasing tip burn risk and reducing marketable yield. This increased variability contrasts with a key advantage of vertical farming, namely the ability to deliver consistent, high-quality produce under controlled growing conditions. Targeted experiments are therefore needed to quantify quality responses under the variable DLI profiles associated with glazing-based daylighting.

In terms of lettuce yield, operating tier 3 with daylighting only (LP_NL) leads to an overall production decrease of 16.7% for the three-tier VF. Conversely, all scenarios combining LPs with LED lighting maintain yield close to the Bench case, with production differences below 3%. Among the hybrid strategies, dimming control results in slightly higher production than the on/off approach, consistent with its ability to better compensate for temporal variability in daylight.

Table 5: Comparative agronomic performance across the scenarios investigated.

| **Scenario** | **Crop [kg year$^{-1}$]** | **WUE [g L$^{-1}$]** | **Total Lighting Energy Consumption [kWh kg$^{-1}$]** | **Harvested Light [MWh year$^{-1}$]** |
|---|---|---|---|---|
| **Bench** | 9,221 | 977.0 | 5.72 | 0 |
| **LP_NL** | 7,682 | 977.0 | 6.02 | 11.1 |
| **LP_Min_200** | 9,023 | 984.0 | 5.92 | 11.1 |
| **LP_Min_250** | 9,179 | 981.0 | 6.02 | 11.1 |
| **LP_Dim** | 9,296 | 993.0 | 5.98 | 11.1 |
| **LP_Dim_IR_98** | 9,247 | 988.0 | 5.39 | 5.0 |
| **LP_Dim_IR_90** | 9,218 | 984.0 | 5.40 | 4.6 |
| **LP_Dim_EC** | 9,205 | 984.0 | 5.86 | 8.2 |
| **GH** | 7,832 | 980.5 | 16.17 | 91.5 |

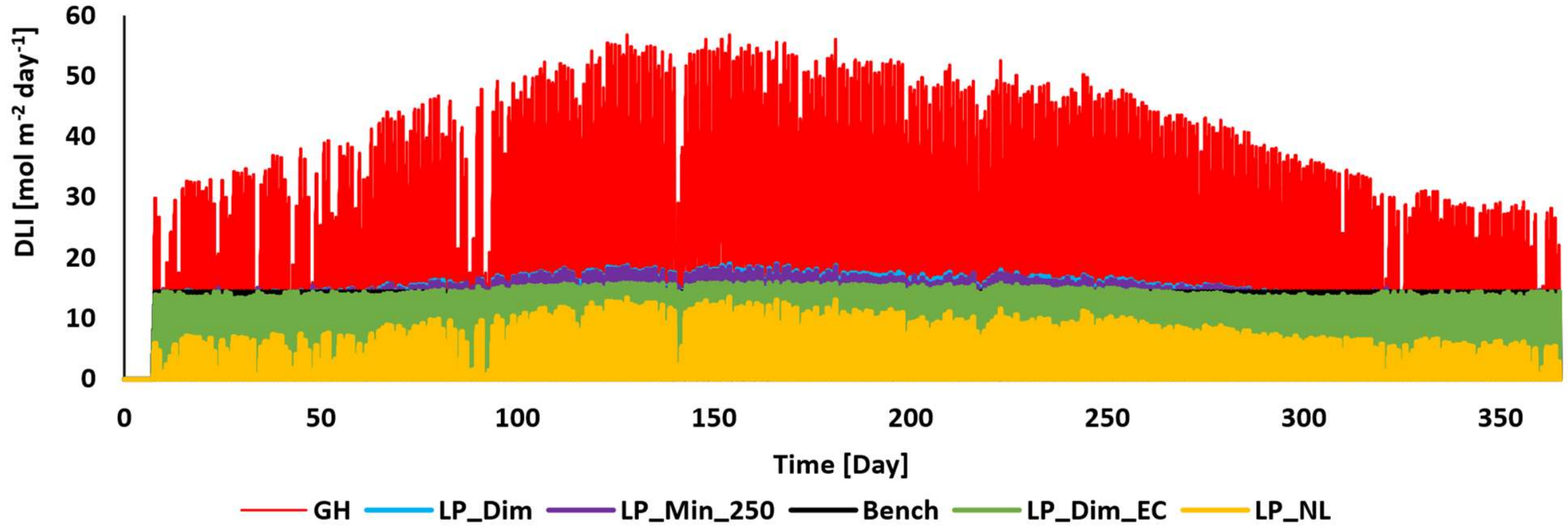

Figure 10: Daily light integral (DLI) profiles for the analysed scenarios. Bench denotes the LED-only baseline. LP_NL denotes the light-pipe, daylight-only case on tier 3. LP_Min_250 denotes the hybrid on/off LED supplementation strategy on tier 3. LP_Dim denotes PWM-dimming LED supplementation on tier 3 to meet a 250 µmol m$^{-2}$ s$^{-1}$ setpoint. LP_Dim_EC denotes PWM dimming combined with variable-transmittance control. GH denotes the greenhouse-like glazing configuration.

### 3.4. Specific Energy Consumption

From an energy perspective, in the benchmark VF, energy use is dominated by lighting and cooling, with lighting slightly higher than cooling, while heating, dehumidification, and post-heating represent minor contributions (Figure *11*). Dehumidification demand remains limited because most evapotranspiration occurs during the light period, when the chiller operates and simultaneously removes both lighting waste heat and water vapor. Owing to the insulated envelope, heating demand is largely confined to the dark period and remains low. Introducing LP-based daylighting consistently reduces lighting electricity demand but increases cooling demand, because a fraction of the incoming solar radiation is not converted into biomass and must ultimately be removed as heat. This lighting-cooling trade-off is a key energetic limitation of daylighting in enclosed VFs. The convective exchange through the LPs also remained a secondary contribution to the chamber energy balance. The maximum convective heat-transfer contribution associated with natural convection inside the pipes was about 1.1 kW, compared with an average cooling load of approximately 8–11 kW across the analyzed scenarios. Therefore, while this term was included for consistency in the energy balance, its influence on the final energetic and economic indicators was limited. Nevertheless, targeting the lighting load remains well motivated, since LED systems account for the largest share of total electricity consumption in vertical farms.

LED integration affects SEC and SEEC through its impact on light delivery efficiency and yield. Compared with the daylight-only scenario (LP_NL), the dimming strategy supplies low-PPFD supplemental light when daylighting is insufficient, improving crop productivity at similar specific consumption. In particular, LP_Dim achieves higher production than daylighting only (LP_NL) at comparable SEEC (about 6.48 kWh $kg^{-1}$) and emerges as one of the most promising configurations by slightly increasing yield relative to Bench while reducing total electricity use by 12%. The on/off supplementation strategy with a lower nominal setpoint (LP_Min_200) further reduces SEC and SEEC by operating tier-3 LEDs at a lower PPFD. This slightly decreases crop production but improves overall light-use efficiency and, consequently, specific energy consumption.

Active optical control via a polymer-dispersed liquid crystal film (LP_Dim_EC) does not provide a net advantage in the analyzed conditions, because the reduction in transmitted useful radiation due to limited optical transmittance outweighs the benefit of suppressing high-PPFD periods. Conversely, UV-IR filtering improves overall performance by removing non-PAR radiation, thereby limiting the increase in cooling demand. With SEEC between 6.31 and 6.37 kWh $kg^{-1}$, LP_Dim_IR_90 and LP_Dim_IR_98 are the most energy-efficient solutions, achieving about 14% lower SEEC than Bench.

For the selected LP_Dim case, an additional sensitivity analysis was carried out by reducing the effective LP optical efficiency at crop level by 10% and 20%. The effect on SEEC remains limited, with increases below 5%. This limited sensitivity is mainly due to the interaction between daylight availability and LED supplementation over the day. In the morning and evening, the reduction in crop-level optical efficiency has only a small effect on electricity use, because the LEDs are already operating and only need to compensate for a slightly larger light deficit. Around midday, especially in summer, the lower daylight input slightly reduces crop productivity. However, this reduction remains modest because these hours are characterized by relatively high PPFD levels, where a moderate decrease in incident light tends to improve light use efficiency and partly compensates for the lower photon input. At the same time, natural light levels remain high enough that LED supplementation is still not required during the central hours of the day, so electricity use is only marginally affected, while cooling demand is also slightly reduced. Since daylighting affects only the upper tier, the impact on whole-farm SEEC is further attenuated. Overall, these results indicate that LP_Dim is reasonably robust to the considered optical uncertainties, with the 10% and 20% derating cases leading to SEEC increases of about 2% and 4%, respectively.

Finally, partial greenhouse-like glazing (GH) reduces lighting electricity demand by 21%, thanks to the larger daylight admission area, but it induces a much larger increase in cooling demand (+183%). The combined effect leads to an overall SEEC of 9.56 kWh $kg^{-1}$, approximately 30% higher than Bench. These results indicate that, despite its simplicity, direct glazing is not suitable for a vertical farm because the penalties in cooling demand and productivity outweigh the lighting savings.

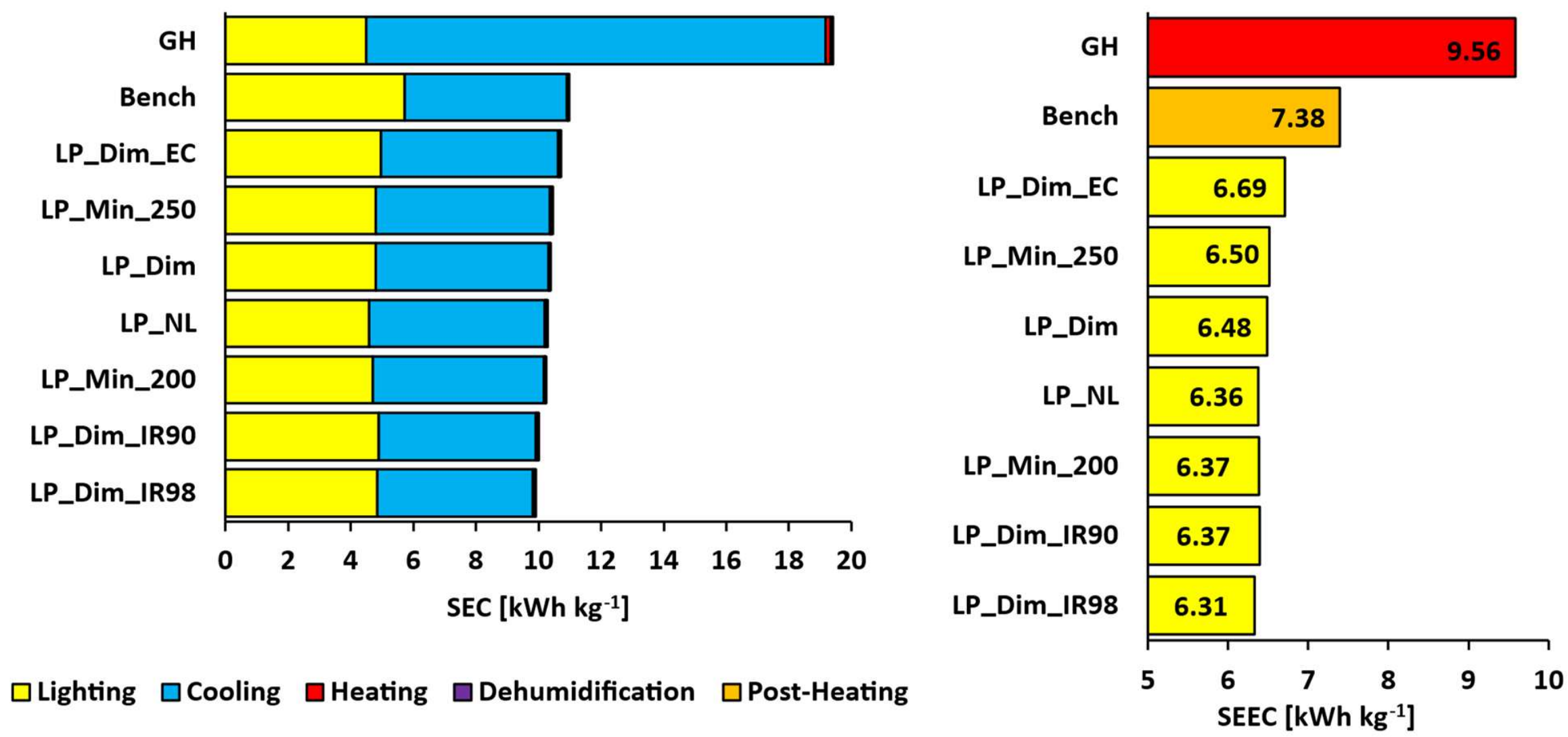

Figure 11: Comparative assessment of SEC and SEEC across the evaluated scenarios.

### 3.5. Power Demand

Figure *12* compares the power-demand profiles of the analyzed scenarios on both a winter and a summer day. In the Bench VF, the load exhibits a quasi-square-wave shape driven by the 16 h photoperiod. Lighting dominates during the light period, while electricity use during the dark period is low because both heating and dedicated dehumidification remain limited. Seasonal effects are primarily associated with cooling performance. In summer, higher ambient temperature reduces the chiller coefficient of performance (COP), resulting in an increase in electricity demand of about 17% relative to winter (see Figure *12*).

The daylight-only configuration (LP_NL) reduces power demand by about 33% during the portions of the photoperiod when daylight is negligible, because tier 3 is not artificially illuminated. As solar irradiance increases, incoming gains raise the cooling requirement and can increase total power demand above the baseline by about 8% in winter and up to 45% at peak in summer. Nevertheless, LP_NL remains below the Bench demand throughout the day because LEDs do not serve one tier.

For the hybrid strategies, power demand matches the Bench during portions of the photoperiod when daylight contribution is negligible and decreases around midday as daylight offsets part of the electric lighting. In particular, LP_Dim_IR_98 exhibits a pronounced midday reduction, with a deeper power trough than LP_NL due to UV-IR filtering, which halves the transmitted solar energy and thereby limits the associated cooling load. The corresponding unfiltered strategies (LP_Dim, LP_Min_250 and LP_Min_200) follow the same temporal pattern but achieve a smaller net reduction in power demand because higher solar gains must be removed by the cooling system. On/off LED supplementation (LP_Min_250 and LP_Min_200) produces a power-demand profile that follows the same overall daily pattern as the dimming strategy (LP_Dim), but with a sharper and time-shifted response. In the morning, the reduction in electricity demand occurs later because LEDs remain fully on until the daylight threshold is reached, while in the evening demand rises earlier as soon as daylight drops below the threshold. This behavior reflects the lower flexibility of on/off control compared with PWM dimming, which can modulate supplemental light more gradually and avoid periods of unnecessary over-illumination.

In both winter and summer, during the central hours of the photoperiod the cooling-related power can exceed the lighting power requirement, despite the heat pump COP, indicating that solar gains dominate the instantaneous load profile under the GH scenario. Moreover, although the increased transmission losses through glazing lead to a slightly higher heating contribution in winter, heating remains a secondary component of the total load.

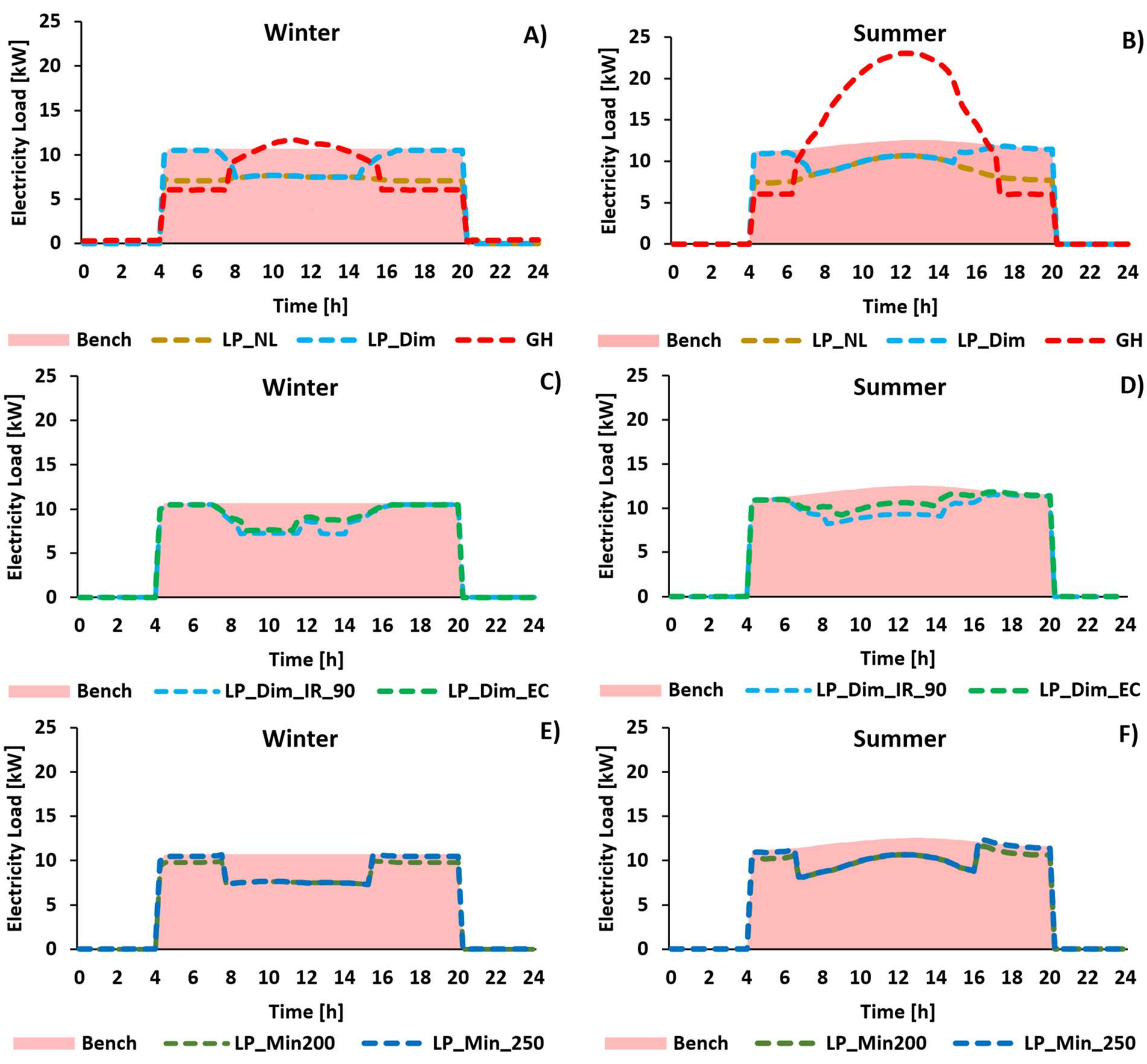

Figure 12: Seasonal comparison of vertical farm electrical power demand across all analyzed scenarios. Bench denotes the LED-only baseline. LP_NL denotes the light-pipe, daylight-only case on tier 3. LP_Min_250 and LP_Min_200 denote hybrid on/off LED supplementation on tier 3 with nominal LED PPFD of 250 and 200 µmol $m^{-2}$ $s^{-1}$, respectively. LP_Dim denotes PWM-dimming LED supplementation on tier 3 to meet the 250 μmol $m^{-2}$ $s^{-1}$ setpoint. LP_Dim_IR_98 and LP_Dim_IR_90 denote LP_Dim with UV–IR filtering, with visible transmittance 0.98 and 0.90. LP_Dim_EC denotes LP_Dim with variable-transmittance control. GH denotes the greenhouse-like glazing configuration.

### 3.6. Economic Assessments

The cost-competitiveness of the proposed daylighting solution is assessed through the light cost (LC), defined as the capital cost per unit photon flux delivered at the crop level, expressed in \$ (μmol $s^{-1}$)$^{-1}$. As a reference, the optical-fiber system presented in [17] reports a luminous output of 9,200 lm under an incident daylight level of 100,000 lx. Assuming the transmitted spectrum lies within the PAR range (400-700 nm), the luminous flux was converted to radiant power using 251 lm $W^{-1}$ and then to photon flux, yielding 167 μmol $s^{-1}$. With a reported system cost of 3,264 \$, the resulting light cost is 19.53 \$ (μmol $s^{-1}$).

Under the same 100,000 lx condition, the LP-based system achieves a delivered photon flux of 24.9 μmol $s^{-1}$ per light pipe. Using the cost assumptions reported in the Appendix [40] to estimate the unit purchase cost of an LP, LC was computed for all analyzed scenarios and is reported in Table *6*. The polymer-dispersed liquid crystal film scenario (LP_Dim_EC) does not provide an economic advantage, as it increases capital costs while reducing transmitted photon flux due to its limited optical transmittance. For the remaining LP scenarios, LC is 15-38% lower than the optical-fiber reference, reflecting the simpler architecture and lower-cost materials. Given the uncertainty in real-world LP purchase and installation costs, a break-even analysis indicates that the LP approach remains more cost-effective than the optical-fiber solution for LP unit cost below 480 \$.

Table 6: Cost comparison between the specific investment cost for the investigated LP systems and the reference optical-fiber system.

| **Scenario** | **Optical Fiber** | **LP_NL** | **LP_Min** | **LP_Dim** | **LP_Dim_IR** | **LP_Dim_EC** | **Unit** |
|---|---|---|---|---|---|---|---|
| **Cost** | 3,264 | 300 | 300 | 300 | 404 | 400 | \$ |
| **PPF** | 167 | 24.9 | 24.9 | 24.9 | 24.4 | 16.0 | μmol $s^{-1}$ |
| **Light Cost** | 19.53 | 12.05 | 12.05 | 12.05 | 16.56 | 25.00 | \$ (μmol $s^{-1}$)$^{-1}$ |

Economic viability is primarily driven by the balance between the additional CAPEX required for LP implementation and the resulting reduction in operating costs, which depends on the tariff. The GH configuration is economically unattractive because the large increase in cooling demand raises total electricity consumption above the Bench scenario, offsetting any reduction in lighting electricity use. The daylight-only strategy (LP_NL) delivers the largest electricity savings, reducing annual electricity demand by 27-29% depending on LED PPE. However, these savings are outweighed by the associated yield penalty, making this configuration economically unfavorable across the considered electricity-price range. Because crop revenue is strongly dependent on marketable yield, any configuration that significantly reduces production becomes economically unattractive, even if it delivers substantial electricity savings.

Hybrid daylight-LEDs integration strategies, such as LP_Dim and LP_Dim_IR_98, achieve more modest electricity savings, on the order of 10-13%, while maintaining or slightly increasing crop production relative to the baseline (Bench). Under current cost assumptions, this level of operating cost reduction is insufficient to compensate for the high upfront investment, leading to long payback times. The performed sensitivity analysis aimed to evaluate the effects of the three main economic drivers on the simple payback time of the proposed solution, namely electricity price, carbon pricing and LP system CAPEX. The results, reported in Figure *13*, show that cost-competitiveness improves mainly under high electricity price scenarios and substantial reduction in LP system cost, while the contribution of carbon pricing remains comparatively limited. Even under high electricity price, the solution remains CAPEX-driven. Assuming an electricity price of 350 \$ $MWh^{-1}$ and a carbon tax of 100 \$ $t^{-1}$, a reduction of 80% in total LP system expenditure would be required to reach a payback time below 10 years, while a reduction of 58% would be required to achieve a payback time below 20 years, assumed here as the system lifetime. If the light pipe purchase cost decreases to 50 \$ $unit^{-1}$, less stringent electricity-price conditions would be sufficient to make the solution cost-competitive. In this analysis, the selling price of vertically farmed lettuce was kept fixed in order to isolate the effect of the main energy-related economic drivers on payback time.

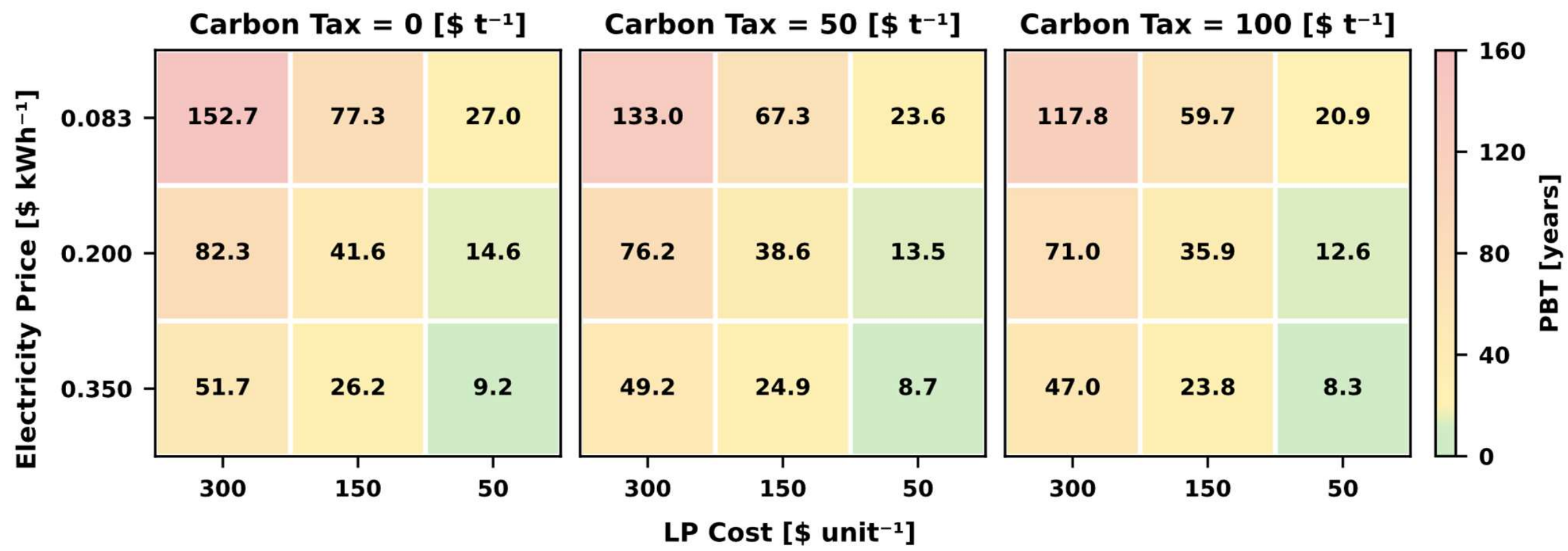

Figure 13: Sensitivity of simple payback time (PBT) to electricity price, carbon tax, and LP unit cost. Values in each cell report PBT in years.

Despite its energy benefit, the UV-IR filter option remains economically unattractive under the assumed costs, as the additional filter CAPEX is not matched by a proportional reduction in electricity consumption.

Under the assumptions adopted for this comparison, the proposed LP concept showed a lower light cost than optical-fiber systems reported in the literature for controlled-environment agriculture, although its technological maturity is still insufficient to ensure cost-effectiveness. At present, no established applications require LP systems at the scale considered here; consequently, unit costs remain high, and the associated CAPEX cannot be offset by the achievable savings in operating costs. However, policy incentives aimed at promoting more sustainable, low-water-demand farming practices, such as vertical farming, could substantially improve the economic feasibility of the daylighting strategy presented here.

The economic results presented here refer to a vertical farm operated at a relatively high PPFD setpoint of 250 µmol $m^{-2}$ $s^{-1}$. Extending the analysis to lower PPFD targets could improve the attractiveness of daylight integration, because the amount of supplemental electric lighting required when daylight is insufficient would decrease. Periods of high daylight availability, particularly in summer, could translate into larger productivity gains relative to the benchmark at these lower setpoints. Ultimately, the reported techno-economic results should be interpreted with reference to compact, low-rise vertical farming systems with roof-constrained daylight collection, while the broader relevance of the study lies in the design trade-offs identified between lighting savings, thermal penalties, and yield preservation.

## 4. Conclusions

This work used a validated transient AGRI-Energy model to quantify the techno-economic impact of integrating a light pipe (LP) daylighting system in a small-scale, three-tier container vertical farm. The LP optical chain was modelled in Tonatiuh to estimate the fraction of daylight effectively delivered to the crop and to distinguish between useful photon flux for growth and additional solar gains affecting the thermal balance. A set of daylighting and integration strategies affecting tier 3 was analyzed and compared against a fully LED benchmark using agronomic, energetic, and economic indicators, including crop productivity, specific electric energy consumption (SEEC), and light cost (LC). A greenhouse-like glazing (GH) case was also included as a simplified daylighting benchmark. Key findings are summarized as follows.

- The proposed LP concept achieved a high optical output, with crop-level LP optical efficiency varying between 45% and 75% depending on solar position. Under identical incident daylight conditions, it showed a lower LC than the selected optical-fiber daylighting reference system.
- All LP-based strategies reduced lighting electricity demand but increased cooling demand because most of the incoming solar energy is not converted into biomass and is rejected as heat. This cooling penalty is the main energetic limitation of daylighting in enclosed VFs and can dominate the instantaneous load during central hours of the photoperiod.
- Daylighting without electric-light integration on the third tier led to a substantial yield penalty. In particular, operating tier 3 with daylighting only (LP_NL) reduced total production by 17%, and despite the associated electricity savings, this scenario was not economically viable. Conversely, hybrid strategies integrating LPs with LEDs through on/off control or dimming maintained yields close to the benchmark, with production differences within a few percent, while also reducing overall electricity requirements.
- The best energy performance was achieved by strategies that limit non-productive radiation while preserving crop output. UV-IR filtering, combined with dimming, yields the lowest SEEC, around 6.32 kWh $kg^{-1}$, corresponding to 14% lower SEEC than the benchmark.
- Although GH admitted more daylight due to the larger admission area, it induced a very large cooling penalty and higher overall SEEC. Under the analyzed conditions, glazing increased SEEC to 9.56 kWh $kg^{-1}$, 30% higher than the benchmark, indicating that the simplest daylighting approach is not energetically favorable for the studied VF.

- In terms of light cost, LP configurations provided a lower cost per delivered photon flux than the optical-fiber reference, with a reduction of about 15-38% depending on the scenario and assumed LP unit cost. However, payback remains strongly constrained by LP-related CAPEX, and scenarios with additional components such as UV-IR filters or variable-transmittance films did not translate their energy benefit into economic viability under the assumed cost structure.

## 5. Appendix and supplementary data

### 5.1 Climatic Data

The container farm analyzed in this study was assumed to be located in Dubai. To quantify transmission losses and assess natural light availability, climate data for outdoor air temperature, DNI, and DHI were collected, and the corresponding profiles are reported in Figure A1. Relative humidity and sky infrared radiation were also included in the analysis, but are not reported here because they play a secondary role in this study.

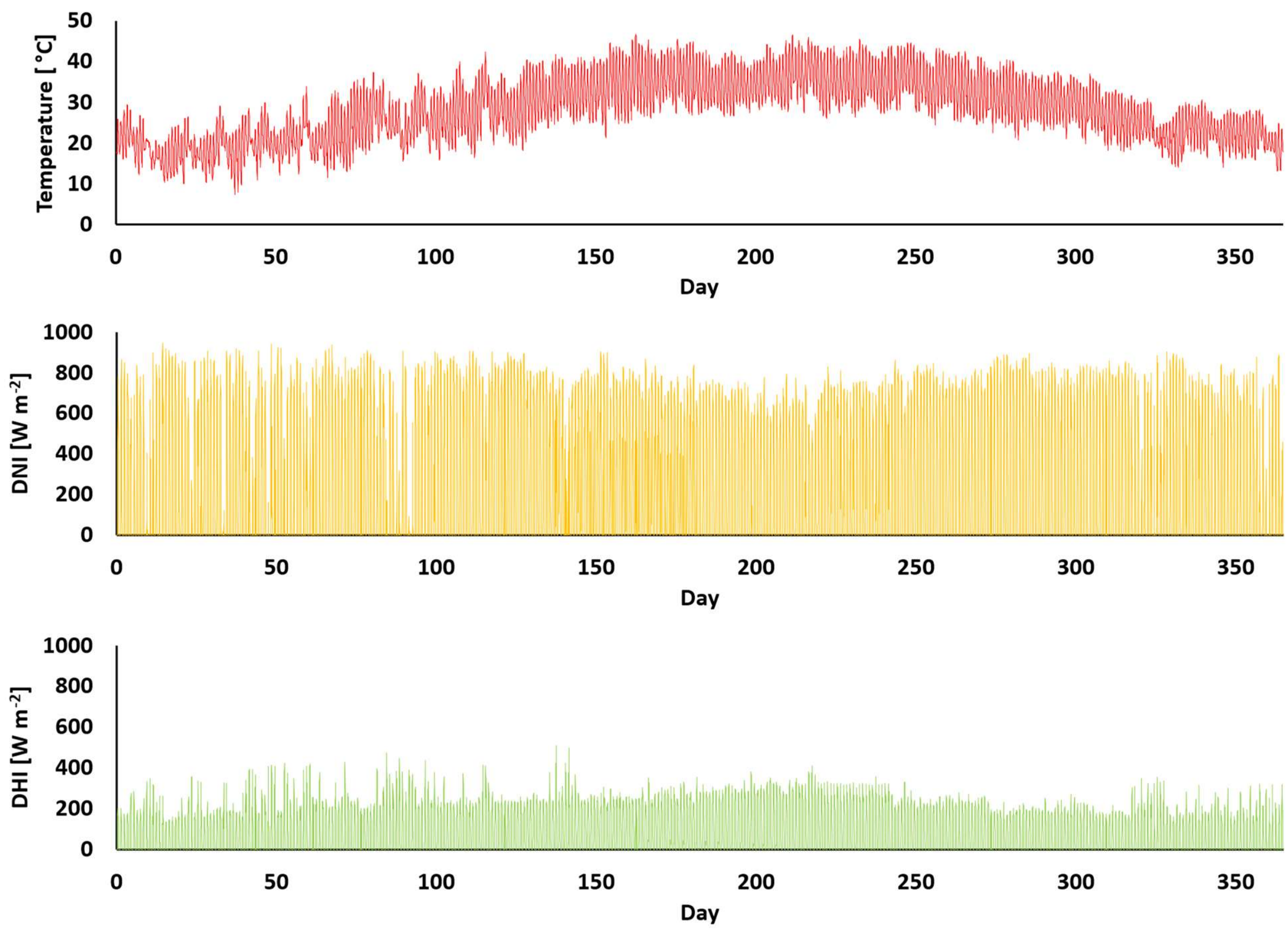

Figure A1: Dubai climate data: outdoor temperature, DNI, and DHI.

### 5.2 Solar Position and Angle of Incidence

The solar position is defined by solar altitude ($\alpha_{sol}$) and azimuth ($\gamma_{sol}$), which depend on geographical location and day of the year. Cooper's equation is used to estimate the solar declination δ (Eq. A1), where n is the day of the year. The Equation of Time EoT (Eq. A2) is used to compute the local solar time $h_{sol}$ from clock time $h_{real}$ (Eq. A3). The auxiliary angle B is computed with Eq. A4, and the hour angle ω is obtained from Eq. A5. Solar altitude and azimuth were then computed using Eqs. A6-A7. The site coordinates were set to latitude (φ) 25° and longitude (ψ) 55°. The reference longitude is 60°, 15*ΔUTC for ΔUTC equal to 4. The resulting solar altitude and azimuth values (with North = 0°) are reported in Figure A2.

$$\delta = 23.45 \cdot \sin(360 \cdot \frac{284+n}{365}) \quad (A1)$$

$$EoT = 9.87 \cdot \sin 2B - 7.53 \cdot \cos B - 1.5 \cdot \sin B \quad (A2)$$

$$h_{sol} = h_{real} + \left(EoT - 4 \cdot \left(\psi - \psi_{ref}\right)\right)/60 \quad (A3)$$

$$B = (n - 81) * \frac{360}{364} \quad (A4)$$

$$\omega = 15 \cdot (h_{sol} - 12) \quad (A5)$$

$$\alpha_{sol} = \arcsin\,(\sin\delta \cdot \sin\varphi + \cos\delta \cdot \cos\varphi \cdot \cos\omega) \quad (A6)$$

$$\gamma_{sol} = \arcsin\,(\frac{\cos\delta \cdot \sin\omega}{\cos\alpha_{sol}}) \quad (A7)$$

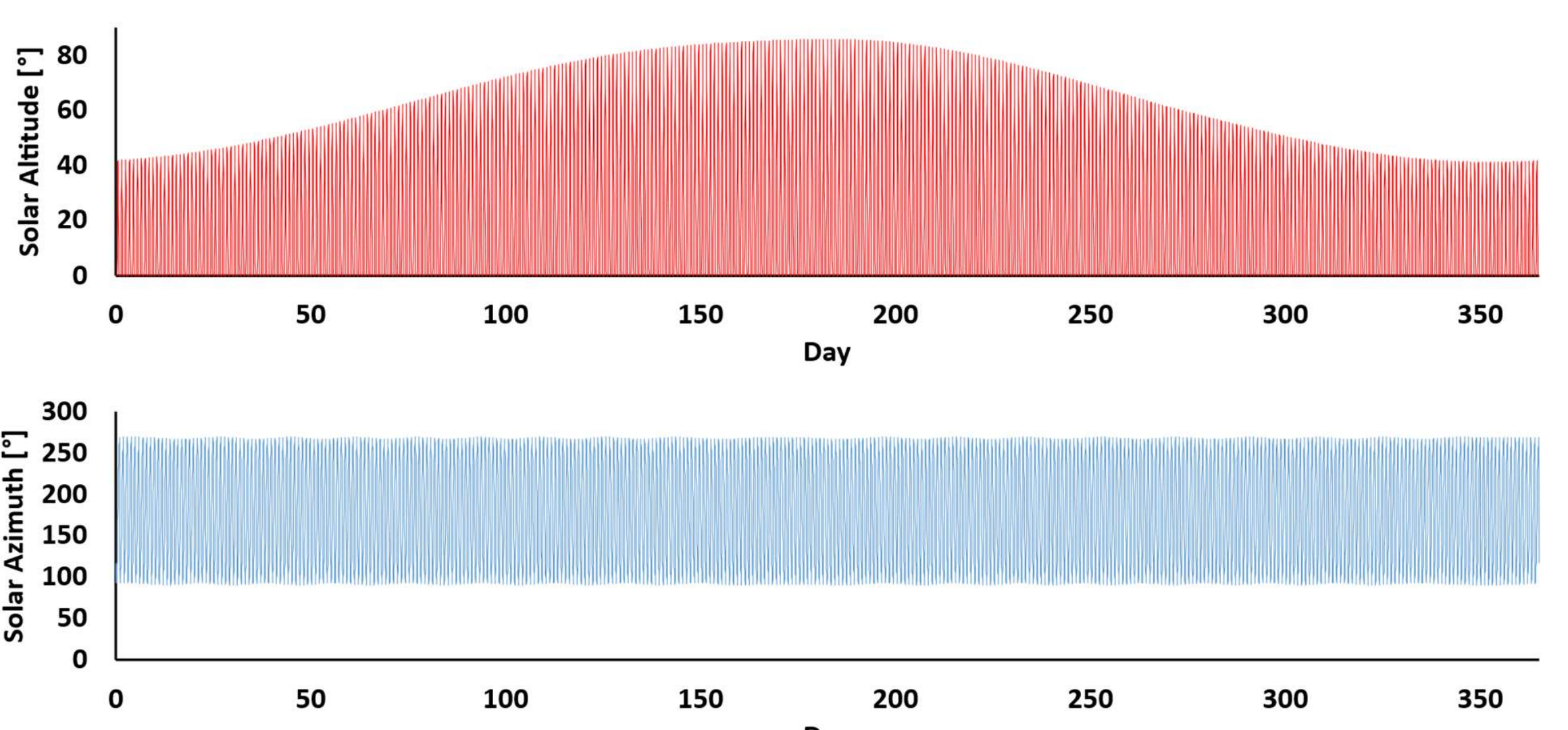

Figure A2: Solar positions in Dubai, including altitude and azimuth angles, referenced to north.

The incidence angle on a horizontal surface was used to evaluate the LP efficiency. As shown in Eq. A8, this angle depends on the solar altitude. The resulting cosine factor is reported in Figure A3.

$$\cos\theta = \sin\alpha_{sol} \quad (A8)$$

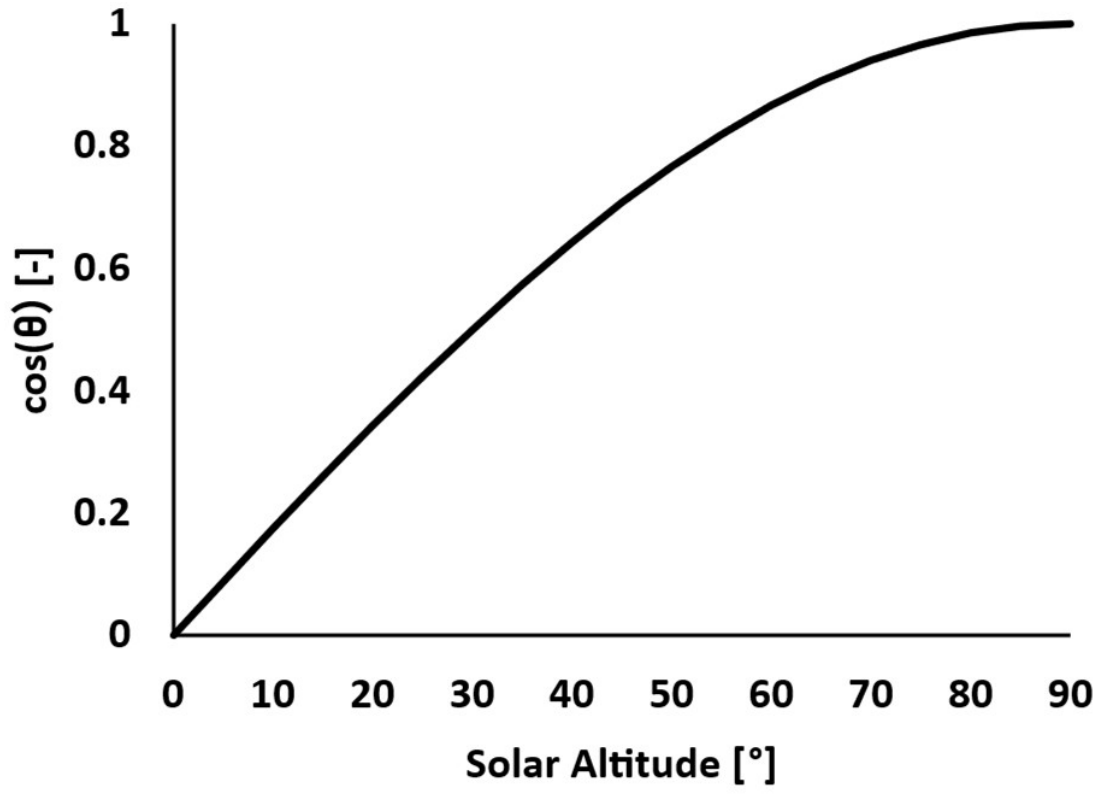

Figure A3: Variation of the incidence factor with solar altitude

### 5.3 CAPEX and OPEX Assumptions

To estimate the specific cost of the LED lighting system, expressed in $ W$^{-1}$, a recent report was adopted, which indicates an installed cost in the range of 25-45 $ ft$^2$ for crops with relatively low lighting requirements [43], such as the lettuce. An average value of 35 $ ft$^2$ was assumed. This was first converted to an area-based cost and then expressed per delivered photon flux assuming a nominal PPFD of 250 µmol m$^{-2}$ s$^{-1}$. The corresponding cost per electrical watt was then obtained by multiplying it by PPE. Cost estimates per LP system are more uncertain due to the high level of customization required for horticultural integration. Nevertheless, commercial light pipes with comparable dimensions exhibit unit costs in the range of 170 to 450 $ LP$^{-1}$, depending on materials and construction. In this study, a representative value within the range was adopted assuming low-cost materials, since the internal aluminium-coated mirror reduces the sensitivity of optical performance to the pipe internal reflectance. To account for the scenario-dependent crop production, revenues were estimated by adopting a constant selling price for vertically farmed lettuce. The PBT of the selected scenarios was computed according to Eqs. A9.

$$PBT = \frac{CAPEX_{scenario} - CAPEX_{Bench}}{c_{el} \cdot \left(E_{el_{Bench}} - E_{el_{scenario}}\right) + c_{crop} \cdot \left(lettuce_{scenario} - lettuce_{Bench}\right)} \quad (A9)$$

Table A1: Economic Assumptions

| Name | Assumed Cost | Unit | Reference |
|---|---|---|---|
| **LED (PPE 3)** | 4.5 | $ W$^{-1}$ | [43] |
| **LED (PPE 2.5)** | 3.75 | $ W$^{-1}$ | [43] |
| **LED (PPE 2)** | 3.0 | $ W$^{-1}$ | [43] |
| **Light Pipe** | 210 | $ LP$^{-1}$ | [40] |
| **Auxiliaries (Mirror, Gears, Motors)** | 90 | $ LP$^{-1}$ | [41] |
| **UV-IR Filter** | 104 | $ LP$^{-1}$ | [17] |
| **Polymer-Dispersed Liquid Film** | 100 | $ LP$^{-1}$ | [44] |
| **HVAC** | 0.65 | $ W$^{-1}$ | [42] |
| **Lettuce Selling Cost** | 7.82 | $ kg$^{-1}$ | [2] |

**5.4 GH Results on PPFD Comparison**

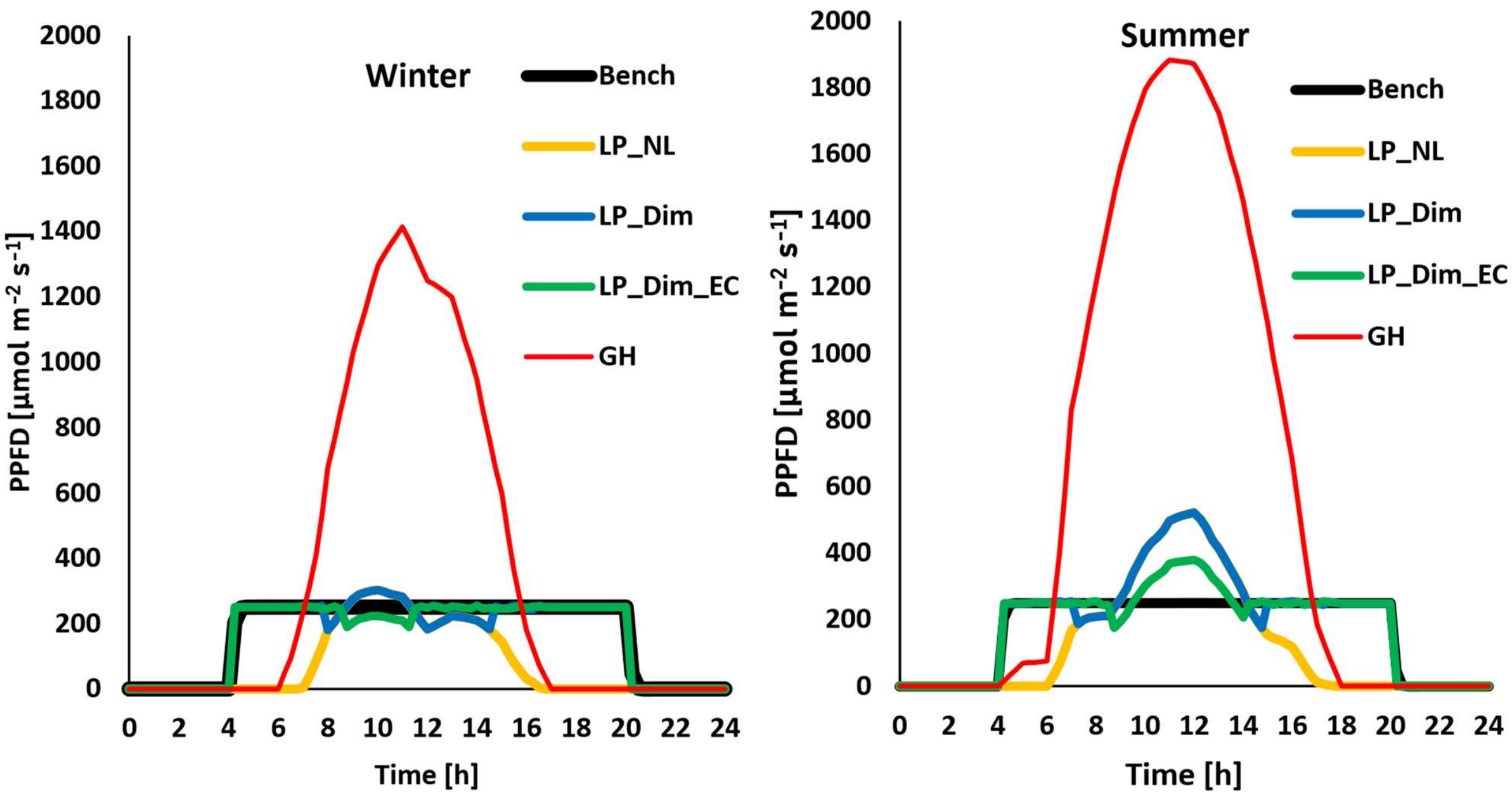

Figure A4: Crop-level PPFD delivered under different LP scenarios and seasonal conditions

**Declaration of generative AI and AI use**

During the preparation of this work, the authors used ChatGPT 4 in order to improve the readability of the manuscript. After using this tool/service, the authors reviewed and edited the content as needed and take full responsibility for the content of the published article.